\documentclass{emulateapj}
\usepackage{psfig,lscape,rotate}                
\begin{document}
\newcommand{\fek}{Fe~K${\rm \alpha} \,$}
\newcommand{\nh}{$N_{H} \,$}

\title{A Simultaneous {\it RXTE} and {\it XMM-Newton} Observation of the 
Broad-Line Radio Galaxy 3C~111}

\shorttitle{{\it RXTE} and {\it XMM} observation of 3C~111}

\author{Karen T. Lewis\altaffilmark{1}, Michael Eracleous\altaffilmark{1},
Mario Gliozzi\altaffilmark{2}, Rita M. Sambruna\altaffilmark{2}, \&
Richard F. Mushotzky\altaffilmark{3}}

\altaffiltext{1}{Department of Astronomy and Astrophysics, The
Pennsylvania State University, 525 Davey Laboratory, University Park,
PA 16802, e-mail: {\tt lewis, mce@astro.psu.edu}}

\altaffiltext{2}{Department of Physics and Astronomy and School of 
Computational Sciences, George Mason University, 4400 University Dr.
Fairfax, VA 22030, e-mail: {\tt mario, rms@physics.gmu.edu}}

\altaffiltext{3}{Laboratory for High Energy Astrophysics, Goddard 
Space Flight Center, Code 660, NASA, Greenbelt, MD 20770, email: {\tt richard@milkyway.gsfc.nasa.gov}}

\begin{abstract} We present the results of simultaneous {\it
XMM-Newton} and {\it RXTE} observations of the Broad-Line Radio Galaxy
3C~111. We find that the Compton reflection bump is extremely weak,
however, broad residuals are clearly present in the spectrum near the
\fek emission line region.  When fitted with a Gaussian emission line,
the feature has an equivalent width of 40--100~eV and full-width at
half maximum of greater than 20,000$\; {\rm km\;s^{-1}}$, however the
exact properties of this weak line are highly dependent upon the
chosen continuum model.  The width of the line suggests an origin in
the inner accretion disk, which is, however, inconsistent with the
lack of Compton reflection.  We find that much of the broad residual
emission can be attributed to continuum curvature. The data are
consistent with a model in which the primary powerlaw continuum is
reprocessed by an accretion disk which is truncated as small
radii. Alternatively, the primary source could be partially covered by
a dense absorber. The latter model is less attractive than the former
because of the small inclination angle of the jet of 3C~111 to the
line of sight. We consider it likely that the curved continuum of the
partial covering model is fortuitously similar to the continuum shape
of the reprocessing model. In both models, the fit is greatly improved
by the addition of an unresolved \fek emission line, which could arise
either in a Compton-thin obscuring torus or dense clouds lying along
the line of sight.  We also find that there are unacceptable residuals
at low energies in the MOS data in particular, which were modeled as a
Gaussian with an energy of $\sim 1.5$~ keV; we attribute these
residuals to calibration uncertainties of the MOS detectors.

\end{abstract}


\section{INTRODUCTION}\label{intro}

In the early 1990s, observations with the {\it Ginga} satellite revealed
that the spectra of many Seyfert 1 galaxies contain an \fek emission
line, with an equivalent width ($EW$) of 100--300~eV, as well as a hard
excess above 10~keV, relative to the simple powerlaw spectrum fitted
over the interval from 2--8~keV \citep*{p90,piro90,mat90,np94}. These
features were readily interpreted as signatures of the reprocessing of the
primary X-ray continuum emission by nearby Compton-thick material,
such as the accretion disk or the obscuring torus \citep*[see,
e.g.][]{lw88, gf91,mpp91,mpps92}. These features are important
diagnostics of the geometry, dynamics, and physical conditions of the
reprocessing medium.

In some Seyfert 1s, most notably MCG --6-30-15 \citep{t95,i96}, {\it
ASCA} observations showed that the \fek line profile had a narrow core
at 6.4~keV and a very broad red wing; \citet{f95} found that the line
profile was best modeled as emission from the inner regions of the
accretion disk. In a sample of 18 Seyfert 1s observed with ASCA,
\citet{n97} found that many had a broad \fek emission line with an
average Gaussian energy dispersion of $\langle\sigma\rangle = 0.43 \pm
0.12$ keV. The averaged profile of the \fek emission lines, as well as
some {\it individual} profiles, were not symmetric however, indicating
that multiple line components (i.e. broad and narrow) were present. In
some objects the line profile had a broad red wing and, like MCG
--6-30-15, were well modeled as emission from an accretion disk.

However, recent observations with {\it XMM-Newton} and have shown that
the picture is much more complex. Broad lines, typically with
equivalent widths of $\sim 100$~eV, were detected in several Seyfert
1s, for example: MCG --6-30-15 \citep{w01}; MCG~--5-23-16
\citep*{d03}; NGC~3516 \citep{t02}; Mrk~766 \citep{p03}; and Ark~120 
\citep{v04}. On the other hand, {\it XMM} has shown narrow \fek lines 
(typically unresolved by EPIC) to be a ubiquitous feature of Seyfert 1
spectra and in fact some Seyfert 1s showed {\it only} a narrow,
neutral \fek line, with $EW \sim 75$~eV \citep[e.g.,][]{r03}. It must be noted,
however, that the upper limits for the equivalent width of a broad
line were sometimes quite generous ($EW \sim 100$~eV), so a broad
component, although not required, was not always ruled out by the
data.

The hard excess from 10--18~keV observed by $Ginga$ is only the
low-energy tail of the Compton reflection bump, a high energy
component which peaks at $\sim 30$~keV and continues up to energies of
100~keV. The strength of the Compton reflection bump is parameterized
by $\Omega/2\pi$, where $\Omega$ is interpreted as the solid angle
subtended by the reprocessing material to the primary X-ray
source. (In the case of a standard accretion disk, $\Omega/2\pi = 1$.)
\citet{np94} measured $\Omega/2\pi$ = 0.5--0.7 using $Ginga$
data, but constraining the properties of the Compton bump well
requires very broad spectral coverage. \citet{g96} combined data from
{\it Exosat}, {\it Ginga}, {\it HEAO-1} and {\it GRO/OSSE} to obtain
an average 1--500~keV spectrum of 7 Seyfert 1s and found that
$\Omega/2\pi = 0.76 \pm 0.15$. More recent observations with {\it
BeppoSAX} \citep{p02,b04} and the {\it Rossi X-ray Timing Explorer}
\citep{l99,wk98} also indicated that $\Omega/2\pi \sim 0.7$, on
average.

Observations with numerous X-ray satellites have shown that these
reprocessing features are stronger in Seyfert 1s than in their
radio-loud counterparts, the Broad-Line Radio Galaxies (BLRGs). The
Compton reflection bump is typically much weaker, with $\Omega/2\pi
\lesssim 0.5$ \citep*[see, e.g.,][]{z95,w98,esm00,zg01,g01a,g01b}. The
\fek line is also weaker, with $EW \lesssim 100$~eV \citep*[see,
e.g.,][]{ehl96,w98,eh98,esm00,g01a,g01b,zg01}. 

Recently, \citet*{bfi04} analyzed simultaneous $XMM$ and $RXTE$
observations of the BLRG 3C~120 and found that $\Omega/2\pi \sim 0.5$
and the \fek line had an $EW \sim$ 50 eV. The line, when fitted with a
Gaussian, had a width $\sigma \sim 0.1$ keV, which is much narrower
than those found in Seyfert 1s \citep{n97}. However, a contribution from a
broad, distorted emission line from the inner accretion disk could not
be ruled out completely. Similar results were obtained by
\citet{o04}, using the {\it XMM} data only.

There are several viable scenarios which could explain the weakness of
the reprocessing features in BLRGs:
\begin{enumerate}

\item{The inner accretion disk in BLRGs might have the form of an ion
torus \citep{rees82}, or other similar radiatively inefficient
accretion flow. As a result, the primary X-ray continuum can only be
reprocessed by either the outer accretion disk or the obscuring torus,
leading to $\Omega/2\pi<0.5$ \citep[see, e.g.][]{esm00}. In this case,
the \fek emission line should be narrow (FWHM $\lesssim 15,000
\; {\rm km \; s^{-1}}$) and produced by Fe atoms which are not highly ionized 
(E$\sim 6.4$~keV).}

\item{\citet*{brf02} suggest that the reprocessing features are weaker
in BLRGs because the accretion disk is highly ionized, rather than
because the geometry of the accretion disk is changed. Reprocessing of
the X-ray emission by ionized media has been studied extensively
\citep[see, e.g.,][]{rf93,z94,nk01}.  These authors find that
reprocessing by a moderately ionized accretion disk results in
numerous low-energy emission and absorption features, due to ionized
species of O, C, and N. However, as the ionization increases further,
the disk becomes a nearly perfect reflector, making the reprocessing
features very weak. In this case, the \fek line should be emitted in
the inner accretion disk and it should be broad, but with $E
\gtrsim 6.7$~keV.}

\item{The weak reprocessing features could be the result of 
a mildly relativistic outflow, as suggested by \citet{w98}. 
This hypothesis is supported by detailed modeling of the effects of
bulk motion on the \fek emission line \citep{rf97} and the Compton
reflection bump \citep{b99}.}

\end{enumerate}

In this paper, we present the results of a simultaneous observation of
the BLRG 3C~111 with {\it XMM} and {\it RXTE}, with the aim of testing
the first two of these hypotheses, which make very clear predictions
for the properties of the \fek emission line. The general properties
of 3C~111, including the results of previous X-ray studies, and the
specific goals of this analysis, with respect to 3C~111, are given in
\S\ref{3c111}. In \S\ref{data}, we describe the data reductions. The
results of the timing and spectral analysis are presented in
\S\ref{timing} and \S\ref{spectral} respectively. We discuss the
implications of these results in \S\ref{discussion} and summarize our
findings in \S\ref{conclusions}. Throughout this paper, we use a WMAP
cosmology \citep[$H_{\rm 0} = 70 {\rm km \; s^{-1} \;Mpc^{-1}},
\Omega_{M} = 0.27, \Omega_{\Lambda} = 0.73$;][]{wmap}.

\section{Properties of 3C~111}\label{3c111}

3C~111 is a nearby ($z$ = 0.0485, d = 210 Mpc) BLRG. The host galaxy
is marginally resolved in the R-band with the {\it Hubble Space
Telescope} and although the morphology of the host is somewhat
uncertain, it is likely to be a small elliptical-type galaxy
\citep{M99}. The radio source has an FR II radio morphology
\citep{fr74} with a single-sided jet \citep{lp84}. The jet exhibits
superluminal motion \citep{vc94}, which along with the apparent size
of the radio lobes \citep{n93}, allows us to place constraints upon
the jet inclination. As described fully in Appendix \ref{inclination},
the jet is inclined at an angle of 21$^{\rm \circ}$--26$^{\rm
\circ}$. If 3C~111 happens to be a rare Giant Radio Galaxy, the
inclination angle could be as small as 10$^{\rm \circ}$ though.

A giant molecular cloud lies along the line of sight to 3C~111 and
some care must be taken to estimate the total Galactic Hydrogen column
density; not only is there a significant contribution from molecular
Hydrogen, but the molecular Hydrogen column density is expected to
vary due to the presence of AU-scale structures in the molecular
cloud. Using the \ion{H}{1} map of \citet*{elw89} and detailed ${\rm
H_{2}CO}$ studies of the foreground molecular cloud by \citet*{mmb93}
and \citet{mm95}, the {\it total} Galactic column density towards
3C~111 is estimated to be 1.2$\times10^{22} \;{\rm cm}^{-2}$. This
value, is expected to vary by several 10$^{21} \;{\rm cm}^{-2}$,
however.

Numerous X-ray satellites have been used to observe 3C~111, most
importantly {\it Ginga} \citep{np94,w98}, {\it ASCA}
\citep*{w98,r98,sem99}, and {\it RXTE} \citep{esm00}, which have the high
energy coverage necessary to study the reprocessing features.  There
are many unanswered questions regarding the X-ray spectral properties
of 3C~111 though, that underscore the difficulties encountered when
analyzing and interpreting the weak reprocessing features in many
BLRGs.

First, it is uncertain whether a reflection component is even
necessary to fit the spectrum of 3C~111. Using {\it RXTE} and {\it
Ginga} data respectively, \citet{esm00} and \citet{w98} found that
$\Omega/2\pi$ is consistent with 0, at the 90\% confidence level. It
is important to place tight constraints upon the continuum emission
since it has important implications for the geometry of the
reprocessing medium. Equally important is the need to robustly fit the
continuum to ensure that the residual \fek emission line can be
properly fitted.  For example, when X-ray spectra of 3C~120 were
fitted with absorbed powerlaw models \citep{r97,g97,sem99}, the \fek
line was found to be very strong and broad ($EW \sim 0.5$ -- 1~keV and
$\sigma \sim 1$~keV). However, when Compton reflection or a broken
powerlaw models were used \citep{w98,esm00,zg01}, the line width and
the equivalent were significantly reduced ($EW \sim 100$~eV and
$\sigma \sim 0.3$~keV.)

Secondly, while the equivalent width of the \fek emission line in
3C~111 was found to be weak ($EW \sim 60^{\;+20}_{\;-10}$~eV) by both
\citet{esm00} and \citet{w98}, the energy and width of the line were
uncertain. \citet{esm00} fitted the line with a Gaussian with a fixed
energy of 6.4 keV and constrained the full-width half-max (FWHM) of
the line to be less than 44,000~${\rm km \;s^{-1}}$, while \citet{w98}
fitted the line with a narrow Gaussian ($\sigma = 0.1$~keV) and found
that $E$~=~$6.7^{\;+ 0.4}_{\;-0.3}$~keV. Furthermore, if the \fek line
originates in the accretion disk, a Gaussian model is a poor
approximation to the true disk line profile and yields misleading
estimates of the line energy and width. The line was marginally
detected in an {\it ASCA} observation \citep{r98} and the line
properties were not well constrained. Thus, neither the origin of the
line (i.e. inner disk vs. outer disk, or an even more distant
reprocessor, such as the torus) nor the ionization state of the
reprocessor are known. In order to evaluate the competing scenarios to
explain the weakness of the reprocessing features of BLRGs presented
in \S\ref{intro}, these parameters must be better constrained.

A simultaneous {\it XMM-Newton} and {\it RXTE} observation of 3C~111
can help address these and other issues. It is important to obtain
simultaneous observations, since the X-ray flux and spectral
parameters of AGNs in general, and BLRGs in particular, are known to
vary on timescales of several days \citep*[e.g.,][]{gse03}. The high
energy sensitivity of {\it RXTE} is critical for accurately fitting
the continuum and detecting the Compton reflection bump, which peaks
at 30 keV. On the other hand, the good spectral resolution and large
collecting area of {\it XMM-Newton} in the 0.4-10 keV range make it
ideal for use in a detailed study of the \fek emission line
properties. Additionally, the spectrum at low energies will be useful
for constraining the ionization of the disk because reprocessing by a
moderately ionized disk leads to numerous emission and absorption
features at low energies \citep{rf93,z94,nk01} which should be
detectable in the {\it XMM-Newton} spectrum, despite the large
absorbing column.

\section{OBSERVATIONS AND DATA REDUCTION}\label{data}

\subsection{{\it XMM-Newton}}\label{xmm}

3C~111 was observed on 14 March 2001 with the European Photon Imaging
Camera (EPIC) and the Reflection Grating Spectrometer (RGS) on-board
the {\it XMM-Netwon} satellite for a duration $\sim 40$~ks. The p-n
data were obtained in Large Window mode and the MOS data were taken in
the Partial Window mode, using the thin filter. The exact exposure
times and count rates for each instrument are given in Table
\ref{exposures}. The data were processed using the {\it XMM-Netwon}
Science Analysis Software (SAS v5.4.1) using the calibration files
released on 29 January 2003. The EPIC data sets (p-n, MOS 1 and MOS 2)
were filtered to remove all flagged events (e.g. events suspected to
be cosmic rays, bad pixels, etc.). There were intense particle flares
during most of the second half of the observation in which the count
rate increased by a factor of 20--100. The flares were successfully
removed using the Good Time Interval tables provided, which also
removed several smaller flares that were present. As a result,
the exposure times were reduced by $\sim 40\%$ (see Table
\ref{exposures}). 3C~111 is quite bright and normally the effect of the
flares might have been adequately corrected for with background
subtraction. However, we noticed that the source count rate actually
{\it decreased} dramatically during the flaring intervals, leading us
to believe that the flares must have been intense enough to saturate
the telemetry, despite the fact that the count rates were well below
the expected threshold for telemetry saturation (the saturation
thresholds are 1150 and 300 ${\rm counts \; s^{-1}}$ for the p-n and
MOS detectors respectively.)  Therefore, the data collected during the
flaring intervals could not be used in any way.  Finally, the EPIC p-n
data were filtered to include only single and double pixel events
(i.e. PATTERN $\le 4$) whereas the MOS data were filtered to also
include triple pixel events (i.e. PATTERN $\le 12$).

The source counts were extracted from a circle with a radius of 44$''$
for the p-n data, for which the fractional encircled energy is greater
than 90\%. The background was extracted from an annulus centered on
the source with an outer radius of 110$''$ and inner radius of 45$''$;
we experimented with several background regions and found little
difference in the results. Since the MOS data were obtained in Small
Window mode, the largest circular region which could be used to
extract the source spectrum had a radius of 39$''$, which encompasses
88\% of the encircled energy. The background could not be extracted
from the same chip, because it was not read out. Instead a circle of
radius 125$''$ was extracted from a neighboring chip. As with the p-n
data, several measurements of the background were found to be
similar. The response matrices (RMFs) and ancillary response functions
(ARFs) were generated with the calibration files released on 29 Jan
2003. When making the ARFs, the source was treated as a point source,
but the encircled energy was modeled as a function of photon energy.

The RGS data were reduced using the SAS routine {\sc rgsproc}, which
automatically filtered the data, traced the 1st and 2nd order
dispersed image of the source, and selected regions to exclude in the
determination of the background spectrum. The RGS data suffered from
the same flares as the EPIC data, however we found that it was
unnecessary to remove the flares; the background subtracted spectra
with and without the flares removed were similar. When the spectra
were extracted, a separate background file was created for each order.

In the preliminary spectral analysis, it became clear that the p-n
data suffered from X-ray loading, which occurred because the frames
used to calculate the offset map (i.e. the zero-energy level) had a
count rate which was too large. The offset map, and thus the gain,
were therefore incorrect on a pixel-by-pixel basis. Additionally,
because the zero-energy level was too high, double-pixel events were
interpreted as single-pixel events of a higher energy. In many cases
X-ray loading is an extreme effect of photon pile-up. However, we note
that based on the count rate of the filtered p-n data, the chance of
photon pile-up is $< 1\%$ both from the estimates from the {\it
XMM-Newton} handbook and our own estimates based on Poisson
statistics. The X-ray loading in this instance likely was the result
of a flare in one or more of the frames used to calculate the offset
map.  At this time, there is no method to reliably correct for X-ray
loading. Thus we were forced to exclude the p-n data from the spectral
analysis.  With the total loss of the p-n data and the 40\% loss of
MOS data, due to flaring, the total count rate was reduced by 70\%
from that anticipated.

The X-ray loading manifested itself in several ways, but was difficult
to diagnose. There were two revealing symptoms of X-ray loading that
we noticed, which distinguished the phenomenon from pile-up. First,
when the p-n and MOS spectra were fitted with simple absorbed
power-law models, there was an absorption feature between 1.8 --
2.1~keV in the p-n data residuals which was absent in the MOS data. We
initially suspected a calibration error in the effective area near the
Si edge, however this feature was absent in other p-n data with
similarly high signal-to-noise ratio (S/N).  Secondly, the results of
the SAS {\sc epatplot} routine indicated that the fraction of single
pixel events was slightly {\it higher} than expected from theory
whereas the fraction of double pixel events was slightly {\it lower},
the opposite of what was expected for photon pile-up. However, this
effect was not dramatic and might have been easily overlooked had
there not been suspicious residuals in the p-n data near 2~keV.

\subsection{{\it Rossi X-ray Timing Explorer}}\label{rxte}

3C~111 was observed with the Proportional Counter Array (PCA) and the
High Energy X-ray Timing Experiment (HEXTE) onboard the {\it Rossi
X-ray Timing Explorer} ({\it RXTE}) satellite from 14--17 March
2001. The exposure time for each instrument is given in
Table~\ref{exposures}. The reduction procedure is described in detail
in \citet{gse03}. Briefly, the PCA and HEXTE data were screened to
exclude events taken when the Earth elevation angle was $\leq 10^{\rm
\circ}$ and the pointing offset from the optical position was $\geq
0.02^{\rm \circ}$. The PCA data were also filtered to include only
events obtained when the satellite was out of the South Atlantic
Anomaly for more than 30 minutes and also those events whose {\sc
ELECTON-0} parameter was $\leq 0.1$.  The PCA background and light
curve were determined with the L7-240 background developed at the {\it
RXTE} Guest Observer Facility (GOF), using the FTOOLS task pcabackset,
v2.1b.  The appropriate response matrices and effective area curves
for the observation epoch were produced using the FTOOLS v.5.1
software package and with the help of the REX script provided by the
{\it RXTE} GOF. Only PCUs 0 and 2 were combined, since PCUs 1,3, 4
were not always turned on. The background applicable to the HEXTE
clusters was obtained during the observation by dithering the
instrument slowly on and off the source.

\section{Timing Analysis}\label{timing}

\begin{figure}
\epsscale{0.85}
\includegraphics*{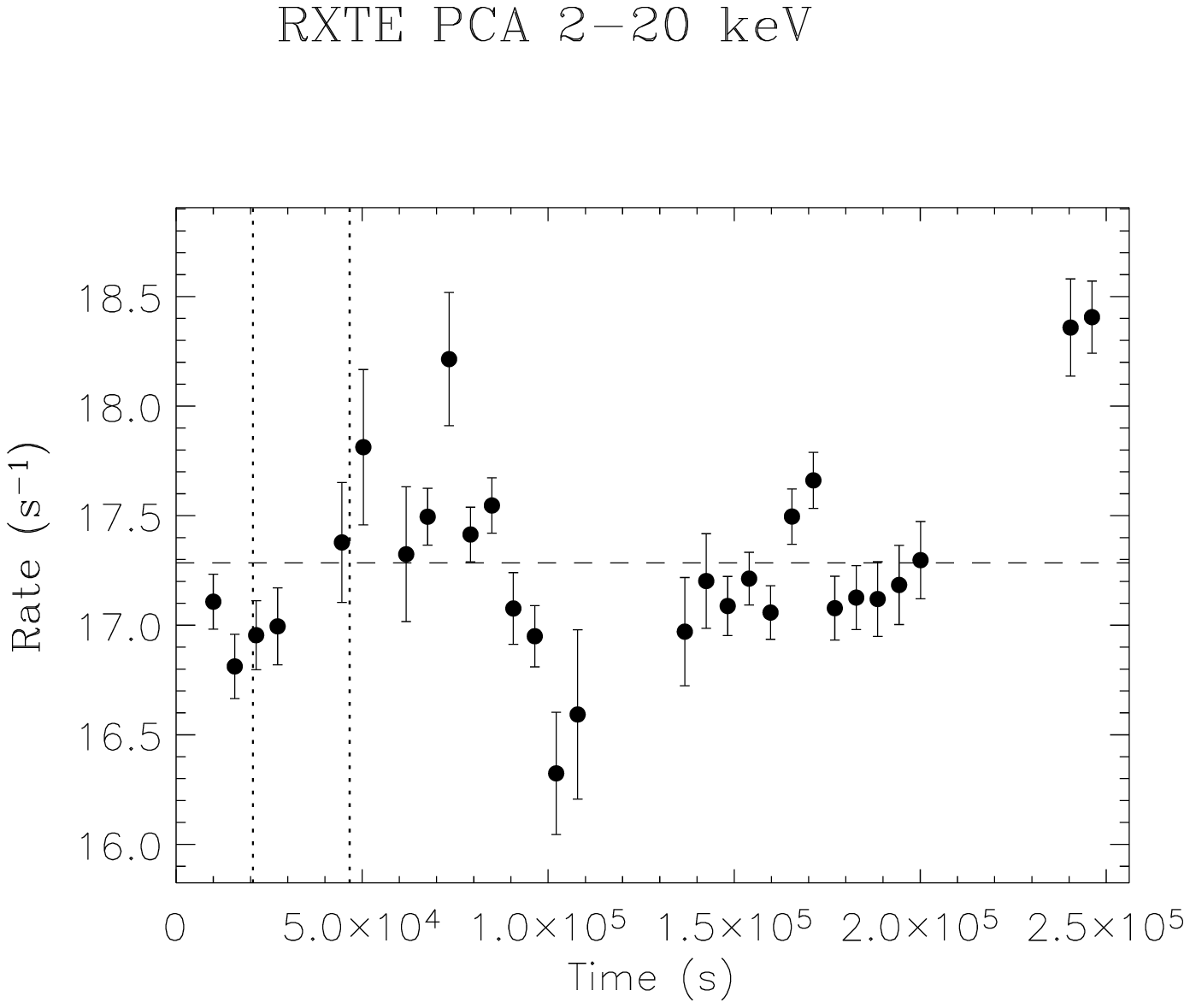}
\caption{\label{lightcurve} The observed RXTE PCA lightcurve in the 2-20 keV energy band.
The dashed vertical lines indicate the interval during which the {\it
XMM} data were taken while the dashed horizontal line indicates the
average count rate over the entire observation.}
\end{figure}

To perform the timing analysis of the {\it XMM} data, the MOS 1 and
MOS 2 data are combined and the p-n data are used, as a slight offset
in the gain should not affect the timing results. The data from
0.2--10 keV are binned in 2000 s intervals.  The mean count rate of
the p-n data is $8.89 \; {\rm s^{-1}}$ and that of MOS 1+2 is $6.98
\; {\rm s^{-1}}$. The count rate is moderately variable, with an
amplitude of 2.7\%, where the amplitude is defined as the difference
between the maximum and minimum count rate, divided by the mean count
rate. However, the variability is significant and the probability that
the count rate is constant is only 2.1\%. The data are consistent with
a monotonic increase in flux during the course of the observation, but
the hardness ratio (defined as the ratio of the flux in the 2--10~keV
band to that in the 0.2--2~keV band) is not significantly variable,
and the probability that it is constant is 43\%.  The 2--20~keV data
from the PCA are also binned in 2000~s intervals and the mean count
rate is $17.29 \, {\rm s^{-1}}$ for 2 PCUs. The variability is consistent
with the {\it XMM} data in the overlapping interval with an amplitude
of 2.5\%. As can be seen in Fig. \ref{lightcurve}, over the entire
observation period, the lightcurve is more variable and the
probability that the count rate is constant is $< 10^{-38}$.  The
largest excursion has an amplitude of 11\% and takes place over a 29
ks time interval.  As with the {\it XMM} data, the hardness ratio
(defined as the ratio of the flux in the 10--20~keV band to that in
the 2--10~keV band) is not highly variable and the probability that
the hardness ratio is constant is 82\%. 

\section{Spectral Analysis}\label{spectral}

In total we have nine separate data sets, obtained simultaneously,
covering the spectral range from 0.4--100~keV: RGS 1 and 2
(0.4--1.65~keV in first order and 0.65--1.65~keV in 2nd order); MOS 1
and 2 (0.4--10.0~keV); PCA (4.0--30.0~keV); and HEXTE clusters 0 and 1
(20.0--100.0~keV). The RGS data are binned such that each bin has a
minimum of 25 counts. The MOS data have an excellent S/N, therefore
they are binned such that there are 2--3 bins per resolution element,
and at least 25 counts per bin. In particular, the region around the
expected \fek line is well sampled. At low energies ($E < 0.7$~keV),
the MOS resolution element was slightly undersampled, because the count
rate is not large. However, this region overlaps with the high
resolution RGS data. The {\it RXTE} data were binned such that there
were at least 20 counts per spectral bin to ensure that the $\chi^{2}$
test was valid.  The spectral analysis is carried out with the XSPEC
{\sc vs 11.3} software package
\citep{xspec}.

The data were obtained simultaneously, so the fit parameters for the
nine data sets are forced to be the same in all models, with the
exception of the overall normalization constants which are allowed to
vary freely to account for cross calibration uncertainties. In
general, there is only a 1\% difference between the normalization
constants for the two MOS data sets, but the PCA normalization
constant is 30\% higher than the MOS value. Since the {\it RXTE}
observations span a longer time period than the {\it XMM}
observations, it is possible that spectral variability will lead to a
systematic difference in model parameters between the {\it XMM} and
{\it RXTE} data, a possibility which we explore below.

All errors and upper limits listed in the Tables and the text
correspond to the 90\% confidence interval for 1 interesting degree of
freedom (d.o.f., i.e. $\Delta\chi^{2} = 2.7$), unless otherwise
stated. In comparing different models, we refer to the chance
probability P$_{\rm c}$, by which we mean the probability that the
improvement in the fit statistic would occur by chance, as determined
by the F-test. In \S\ref{cont_sec}, we describe fits to the continuum
using various models, excluding the energy interval where the \fek
line is expected.  The residual \fek line from several continuum fits
is modeled in \S\ref{line_sec}. Then in \S\ref{combo_sec}, we attempt
to fit the combined continuum and \fek emission simultaneously and
self-consistently.

\subsection{Continuum Models}\label{cont_sec}

We fit the continuum using several different models, excluding the
interval from 4.5--7.5~keV, where an \fek line is expected to be
located. The fit parameters for all models are listed in Table
\ref{cont_param} and discussed below. In all models, we include 
Galactic photoelectic absorption using the cross-sections of
\citet{wabs}, allowing the column density to be a free parameter because
the total Galactic Hydrogen column density along the line of sight is
uncertain. Throughout, we use the solar abundance pattern of
\citet{ae82} and in all instances the fitted column density is $\sim
8\times 10^{21}\;{\rm cm}^{-2}$, which is consistent with the Galactic
Hydrogen column density towards 3C~111.\footnotemark\footnotetext{A
similarly good fit is obtained using the abundance pattern of
\citet{ag89}, but only if the Oxygen abundance is reduced by $\sim$ 50\%. 
In this case, \nh $\sim 10^{21}\;{\rm cm}^{-2}$.}  Therefore an
additional absorber at the redshift of 3C~111 is not warranted,
although absorption within the host galaxy or the AGN itself cannot be
ruled out.

\begin{figure}
\epsscale{0.8}
\plotone{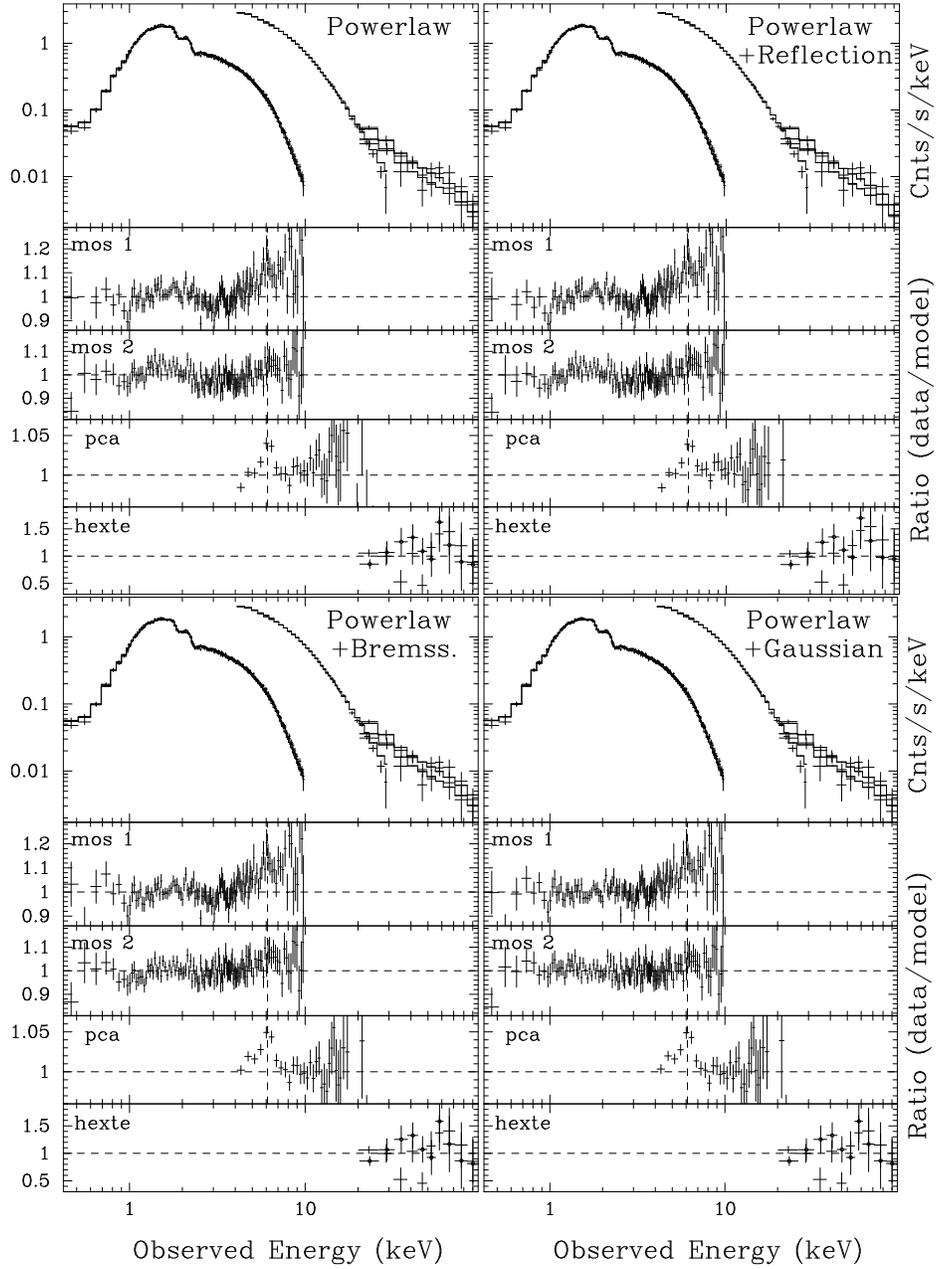}
\caption{\label{spectra} {\it Top panel -} Observed 0.4--100~keV
spectrum with the different continuum models described in
\S\ref{cont_sec} overlaid (the fits excluded the data from 4.5--7.5
keV): Powerlaw (top left), Powerlaw + Compton Reflection (top right),
Powerlaw + Bremsstrahlung (bottom left), and Powerlaw + Soft Gaussian
(bottom right) models The RGS data have been excluded and the MOS have
been binned for clarity. {\it Lower panels -} Ratio of the data to the
model for, top to bottom, MOS 1, MOS 2, PCA, and HEXTE (cluster 0 -
crosses, cluster 1 - filled circles).  The position of 6.4~keV \fek
emission line at the redshift of 3C~111 is indicated by the vertical
dashed line. The line is clearly seen in the PCA data, but is less
obvious in the MOS 1 and MOS 2 data.}
\end{figure}

\begin{figure}
\epsscale{0.7}
\rotatebox{-90}{\plotone{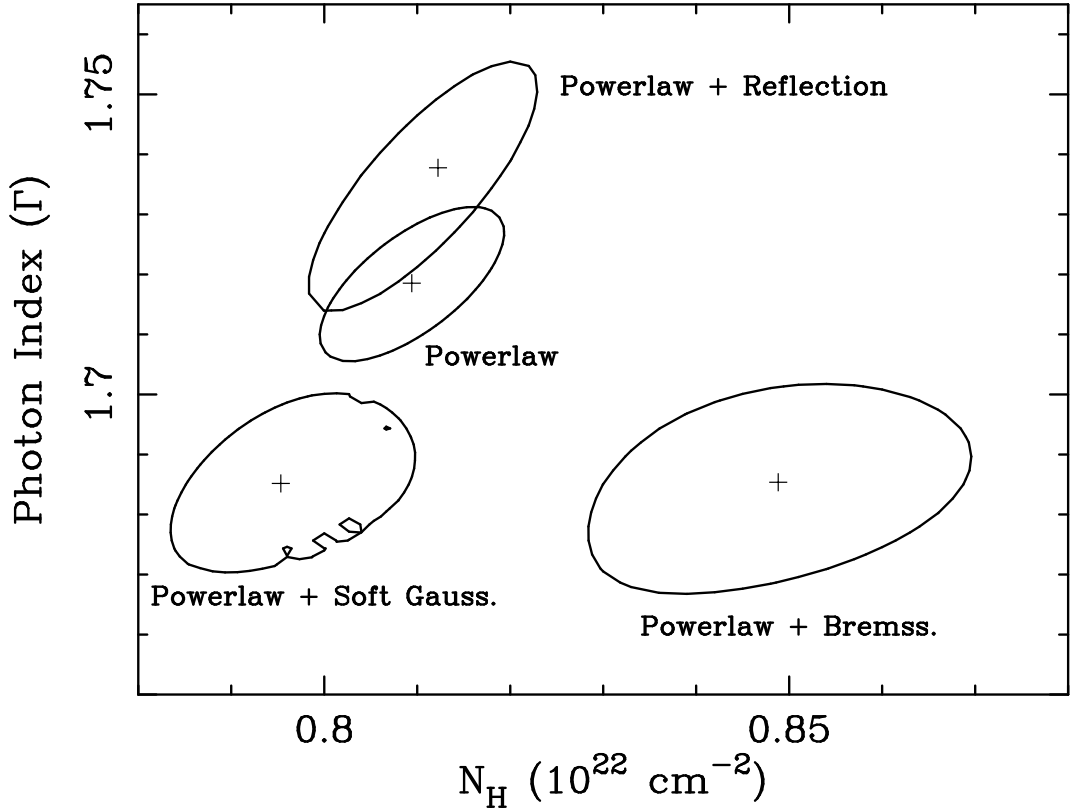}}
\caption{\label{nh_gamma} Confidence contours (90\%) in the
\nh -- $\Gamma$ plane for four different continuum models: Powerlaw
(solid); Powerlaw + Compton Reflection (dotted); Powerlaw +
Bremsstrahlung (dashed); and Powerlaw + Soft Gaussian (dash-dot). The
best-fit powerlaw index for the Powerlaw + Bremsstrahlung and Powerlaw
+ Soft Gaussian models is significantly steeper than that found for
the Powerlaw and Powerlaw + Compton Reflection models.}
\end{figure}

\begin{description}

\item{{\it Powerlaw} (Model \#1): -- The data are first fitted with an absorbed
powerlaw model, whose free parameters are the column density (\nh),
the photon index ($\Gamma$), and the overall normalization
constant. The spectral model and residuals are shown in
Fig. \ref{spectra} and the 90\% confidence contour in \nh and $\Gamma$
is shown in Fig. \ref{nh_gamma}.}

\end{description}

We use this simple model to test the assumption that there are no
systematic differences between the {\it XMM} and {\it RXTE} data
sets. Because $\Gamma$ and \nh are only loosely constrained by the
RGS and HEXTE data, we do not include those data sets in this test.
We allow \nh and $\Gamma$ for MOS 1 and MOS 2 and $\Gamma$ for the
PCA data to vary independently, but we set \nh for the PCA data equal
to that of MOS 1 because \nh is unconstrained by the PCA data
alone. The MOS 1 and 2 data sets yield consistent values of \nh, but
there are discrepancies in $\Gamma$, with $\Gamma_{\rm MOS 1} = 1.74
\pm 0.02$, $\Gamma_{\rm MOS 2} = 1.78^{+0.03}_{\;-0.02}$ and $\Gamma_{\rm
PCA}= 1.69^{+0.02}_{\;-0.01}$. We find that this discrepancy is {\it
not} the result of the spectrum, as whole, becoming harder during the
longer {\it RXTE} observation; fits to several temporal subsets of the
PCA data are best-fit with the same value of $\Gamma$. When the MOS 1,
MOS 2, and PCA data are fitted above 3 keV (excluding the interval
between 4.5--7.5 keV), we find that $\Gamma_{\rm MOS 1} =
1.57^{\;+0.09}_{\;-0.07}$, $\Gamma_{\rm MOS 2} = 1.68 \pm 0.08$ and
$\Gamma_{\rm PCA}= 1.70^{\;+0.02}_{\;-0.03}$. Thus it appears that the
discrepancy in the powerlaw index is an indication that the spectrum
hardens at high energies. However, it is clear that the photon indices
inferred from the MOS 1 and PCA data are inconsistent, although the
MOS 2 and PCA data are in excellent agreement. This can be seen in Fig.
\ref{spectra}: the MOS 1 data show clear positive residuals above the
Fe K$\alpha$ line region that are not present in the MOS 2 data.
Throughout the rest of this paper, we will assume that the {\it XMM}
and {\it RXTE} data can be fitted with the same set of parameter
values and allow only the normalization constants to vary
independently, however the effects of allowing the MOS 1 data to have
an independent photon index will also be considered.

\begin{figure}
\epsscale{0.7}
\rotatebox{-90}{\plotone{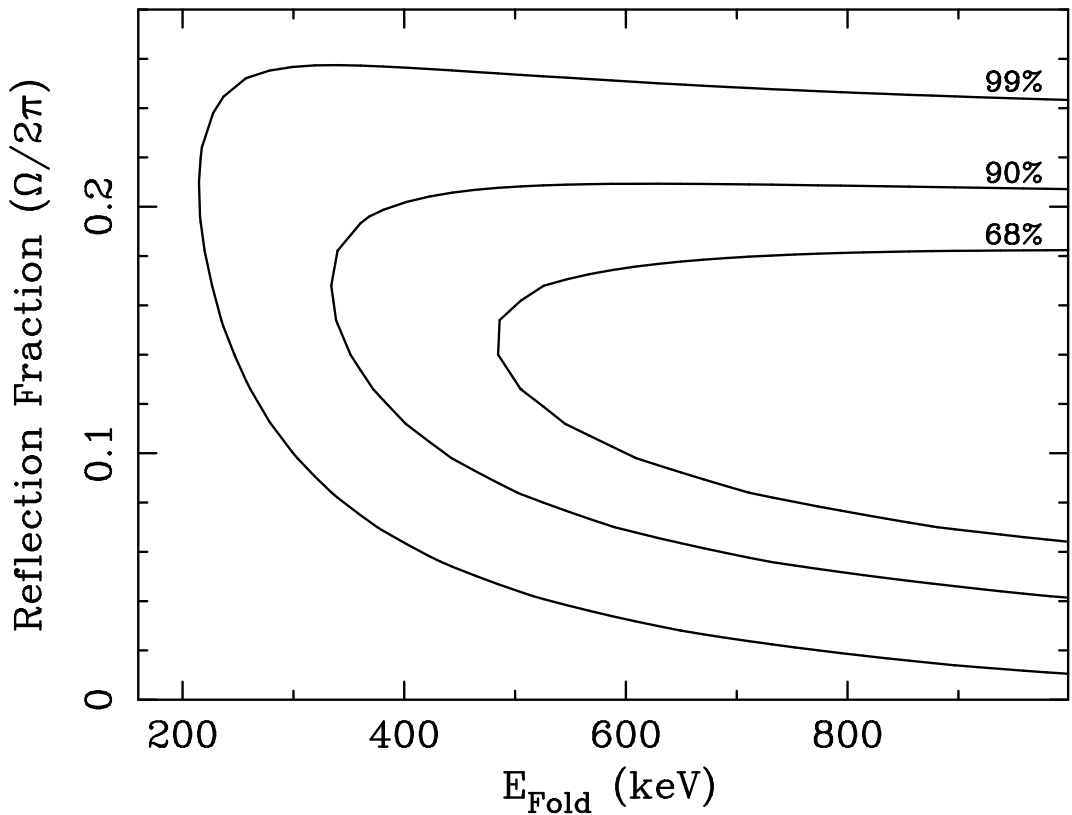}}
\caption{\label{efold_refl} Confidence contours 
in the E$_{\rm fold}$ -- $\Omega/2\pi$ plane for the Powerlaw + 
Compton Reflection model. The inclination was fixed to 
26$^{\rm \circ}$. The reflection fraction is small, but
non-zero.}

\end{figure}

\begin{description}

\item{{\it Broken Powerlaw} (Model \#2): -- The next simplest model is
an absorbed broken powerlaw, whose free parameters are the column
density, two powerlaw indices ($\Gamma_{1}$ and $\Gamma_{2}$), the
breaking energy ($E_{\rm b}$), and the normalization constant.  The
fit is greatly improved, with a chance probability of
$\sim 10^{-16}$. The small breaking energy, $E_{\rm b} = 2.9 \pm 0.2$,
suggests that the addition of a low energy component might improve the
fit greatly. However, this extra component also increases the
complexity of the model; before including it, we first investigate the
effects of including Compton reflection from neutral and ionized
media.}

\item{{\it Powerlaw + Compton Reflection from a neutral medium}
(Model \#3): -- We next fit the data with a model which includes
powerlaw emission from the primary X-ray source as well as the
continuum from X-rays reprocessed by a cold, dense slab of material.
This model, implemented with the {\sc pexrav} routine in {\sc XSPEC}
\citep{mz95}, does not include \fek line emission nor 
does it include the gravitational effects which would arise if the
reprocessing material is near the black hole. The free parameters
are: \nh, $\Gamma$, the folding energy of the primary
X-ray powerlaw continuum ($E_{\rm fold}$), the cosine of the
inclination angle (cos$\;i < 0.95$ or $i > 18^{\rm
\circ}$), and the solid angle subtended by the disk to the primary X-ray
source ($\Omega/2\pi$). Like \citet{esm00}, we find that the elemental
abundances do not greatly affect the fit and we keep them fixed to the
solar values. As described in Appendix \ref{inclination}, the
inclination angle can be further constrained to be less than 26$^{\rm
\circ}$. As $E_{\rm fold}$ increases above 850 keV, the improvement in
the fit is negligible, so we restrict $E_{\rm fold} < 1000$~keV. With
these additional constraints in place, we find that there is a
marginal improvement in the fit as compared to the simple powerlaw
(P$_{\rm c}$ = 17\%). The spectral model and residuals are shown in
Fig. \ref{spectra} and the 90\% contour in
\nh--$\Gamma$ space is included in Fig. \ref{nh_gamma}.
As shown by the $E_{\rm fold}$--$\Omega/2\pi$ confidence contours,
(Fig. \ref{efold_refl}), $\Omega/2\pi$ is small, but non-zero.  We note
that $\Omega/2\pi$ is insensitive to the inclination angle, within the
narrow range of allowed values.}

\item{{\it Powerlaw + Compton reflection from an ionized medium} 
(Model \#4): -- As discussed in \S\ref{intro}, the ionization state of the
disk is important in determining the strength of the reflection
features, so we now consider reflection from an ionized disk, as
implemented with the {\sc pexriv} routine in {\sc XSPEC}
\citep{mz95,d92}. The additional free parameters in this model are
the disk temperature, T ($<10^{6} \;$K) and the ionization parameter,
$\xi=4\pi\;F_{ion}/n$, ($\xi < 5000 {\rm erg\;cm\;s^{-1}}$) where
$F_{ion}$ is the incident 5--20~keV flux and $n$ is the density of the
reflecting medium. The improvement in the fit, as compared to a
neutral reflection model, is negligible (P$_{\rm c} = 37.6\%$) and we
found that $\xi < 20 {\rm erg\;cm\;s^{-1}}$ and $\Omega/2\pi = 0.13
\pm 0.06$.}

\end{description}

When the above models are fitted to the data, there are unacceptable
residuals at low energies (See Fig. \ref{spectra}). Furthermore, the
results of the broken powerlaw fit indicate that the addition of a
soft component is likely to improve the continuum fit
significantly. Thus, we now add a variety of low energy components to
a powerlaw continuum in an attempt to improve the fit. 

\begin{description}

\item{{\it Powerlaw + Thermal Bremsstrahlung} (Model \#5): --
We first add a thermal Bremsstrahlung component to the absorbed
powerlaw model, leading to a significant improvement in the fit
compared to the powerlaw model ($\Delta\chi^{2}$ = 67 for three extra
d.o.f.; P$_{\rm c} \sim 10^{-13}$). The thermal component has a
temperature of $\sim 1$~keV and is found to contribute 10\% and 1\% of
the flux in the 0.5--2~keV and 2--10~keV bands, respectively. The
spectral model and residuals are shown in Fig. \ref{spectra} and the
90\% contour in \nh--$\Gamma$ space are included in
Fig. \ref{nh_gamma}. We note that a $\sim$ 0.3 keV blackbody component
also provides a good fit. Even with the addition of the Bremsstrahlung
or blackbody component however, some low-energy residuals remain.}

\item{{\it Powerlaw + Soft Gaussian} (Model \#6): --
It is also possible that the low-energy residuals are the result of
remaining problems in the calibration of the MOS effective area
curves.  Therefore we also attempt to model the residuals with
combinations of emission lines and absorption lines and edges. We
find that a single Gaussian emission line with $E \sim 1.5$~keV,
$\sigma \sim 0.5$~keV, and a flux of $\sim 10^{-12}\;{\rm
erg\;cm^{-2}\;s^{-1}}$ is quite effective in eliminating the majority
of the low energy residuals ($\Delta\chi^{2}$ = 89 for three extra
d.o.f; P$_{\rm c} \sim 10^{-18}$). The spectral model and residuals and
the 90\% contour in \nh--$\Gamma$ are shown in Figs. \ref{spectra} and
\ref{nh_gamma}, respectively.}

\end{description}

Whether a $\sim 1$~keV Bremsstrahlung, a $\sim$ 0.3 keV blackbody, or
a Gaussian emission line is added, the best-fit powerlaw photon index
decreases to the value we find from fitting only the PCA data ($\Gamma$ =
1.69). Therefore, it is not surprising that when these components are
added to the models which include reprocessed emission, no Compton
reflection is necessary and $\Omega/2\pi < 0.05$. We find that if we
exclude the data from 1--3~keV, where the low-energy residuals are
the most extreme, we find again that $\Gamma$ = 1.69; it appears that
the addition of a low energy component is truly necessary to obtain a
reliable description of the 3--100~keV spectrum. The origin of this
component is discussed further in \S\ref{soft}.

In summary, the data are best fitted with a simple Powerlaw (with
\nh$\approx 8 \times 10^{21} \; {\rm cm^{-2}}$ and $\Gamma \approx 1.7$) with 
little or no contribution from Compton reflection of the primary
powerlaw emission. This is in marked contrast with Seyfert galaxies,
in which $\langle\Omega/2\pi\rangle$ = 0.7 (see
\S\ref{intro}). Throughout the remainder of this paper, a soft
Gaussian component is included in {\it all} models because it provides
the best statistical fit, however the results we obtain using a
thermal Bremsstrahlung (or blackbody) component are similar.

\subsection{Models for the \fek Line}\label{line_sec}

In this section, we study the profile of the residual \fek emission
from the Powerlaw + Soft Gaussian and Powerlaw + Compton Reflection
models (\#6 and \#3 respectively). Before adding the line component,
all data outside the 3.5 -- 8.5 keV energy interval are excluded and
the continuum parameters, including the normalization constants, are
also frozen. The emission line is fitted by either a Gaussian line or
a relativistically broadened line from an accretion disk, implemented
with the {\sc diskline} routine
\citep[see][]{f89}. The Gaussian parameters are the line energy ($E$), the
energy dispersion ($\sigma$), and the line flux.  The {\sc diskline}
model has the following parameters: the inner and outer radius of the
line-emitting region ($R_{in}$ and $R_{out}$, where $R$ is expressed
in units of the gravitational radius, $r_{g} = GM_{\bullet}/c^{2},
{\rm with} \, M_{\bullet}$ the mass of the black hole), the disk
inclination ($i$), the slope $\beta$ of the emissivity pattern,
($\epsilon \propto r^{\beta}$), the energy of the emission line ($E$),
and the line flux.  Note that the diskline model does not include the
effects of the Comptonization of the \fek line as it emerges from the
disk, which asymmetrically broadens it further. The line parameters
for all three data sets are forced to be the same, with the exception
of the normalization constants, which are linked together such that
the equivalent width of the line was the same for all three.

\begin{figure}
\epsscale{0.7}
\plotone{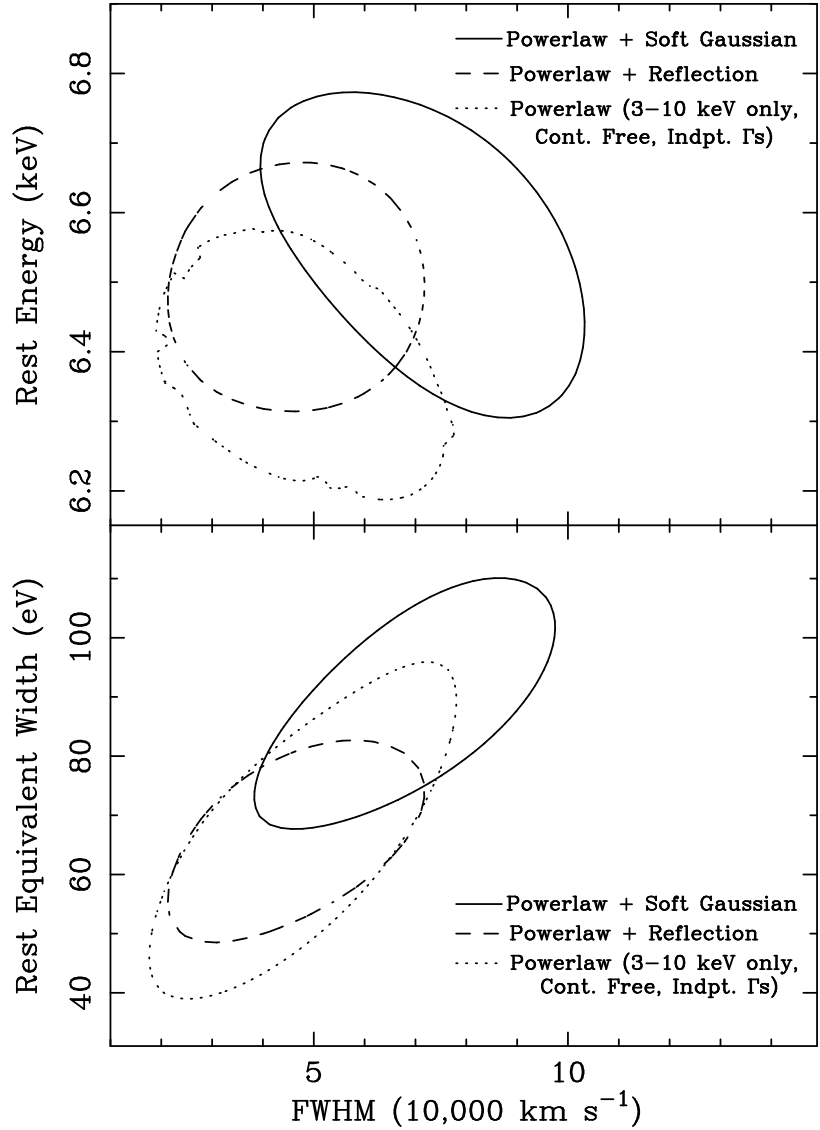}
\caption{\label{gauss} Confidence contours (90\%) for the Gaussian fit
to the \fek emission line to the residuals from the Powerlaw + soft
Gaussian continuum fit (model \#6, solid). For comparison, we show the
contours for the Gaussian fit to the residuals from the Powerlaw +
Compton Reflection fit (model \#3, dashed) as well as the simultaneous
fit to \fek and the continuum over the 3--10 keV interval (dotted, see
\S\ref{line_sec}). The equivalent width is defined to be the line flux
divided by the specific continuum flux at the rest energy of the
emission line, E.}
\end{figure}

\begin{figure}
\epsscale{0.75}
\plotone{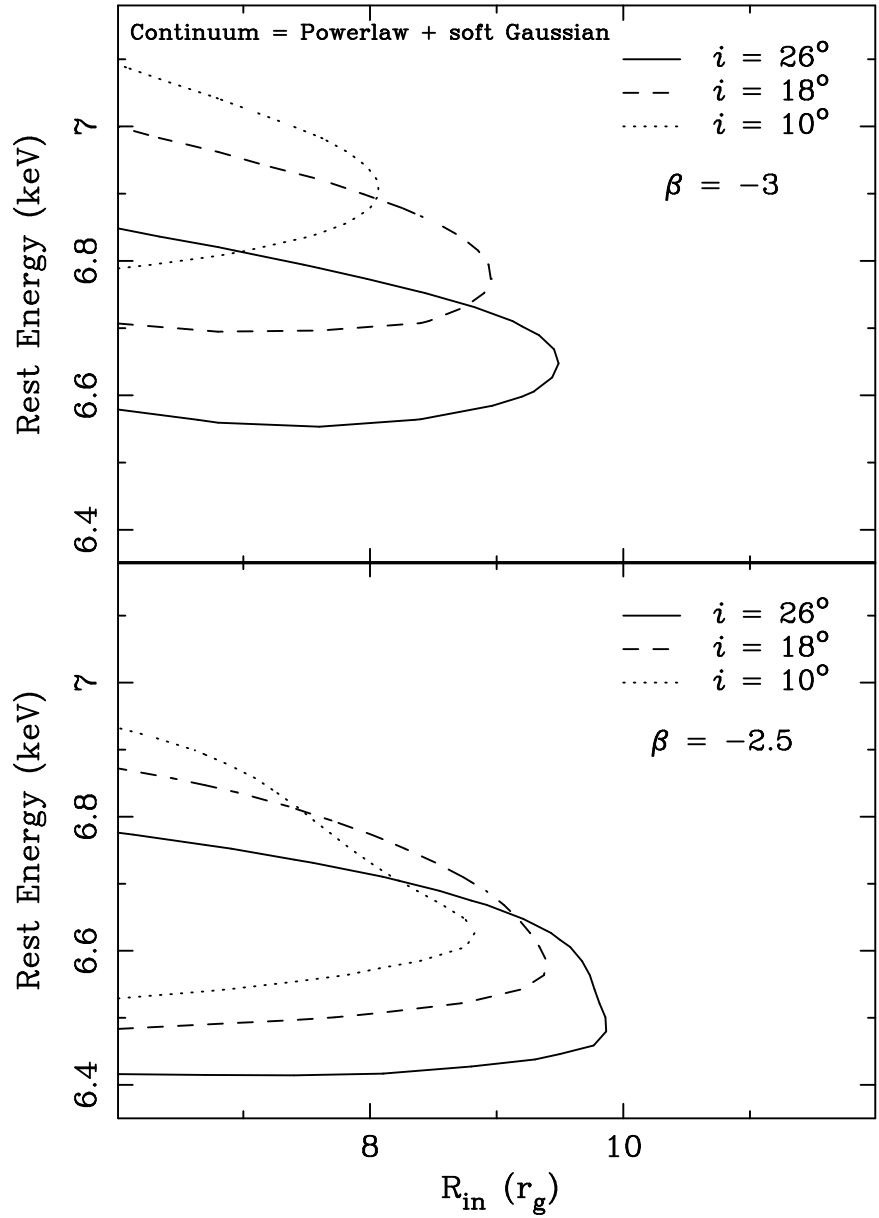}
\caption{\label{diskline} Confidence contours (90\%) for the diskline
fit to the \fek emission line in $R_{in}$--$E$ space, using the
Powerlaw + soft Gaussian continuum model (Model \#6). The shape of the
line profile is sensitive to the inclination $i$ and the powerlaw
emissivity index, $\beta$, thus we present a set of $R_{in}$--$E$
contours for $i$ = 26$^{\rm \circ}$ (solid), 18$^{\rm \circ}$ (dashed), and 10$^{\rm \circ}$ (dotted)
using two different values of $\beta$.}
\end{figure}

\begin{figure}
\epsscale{0.6}
\plotone{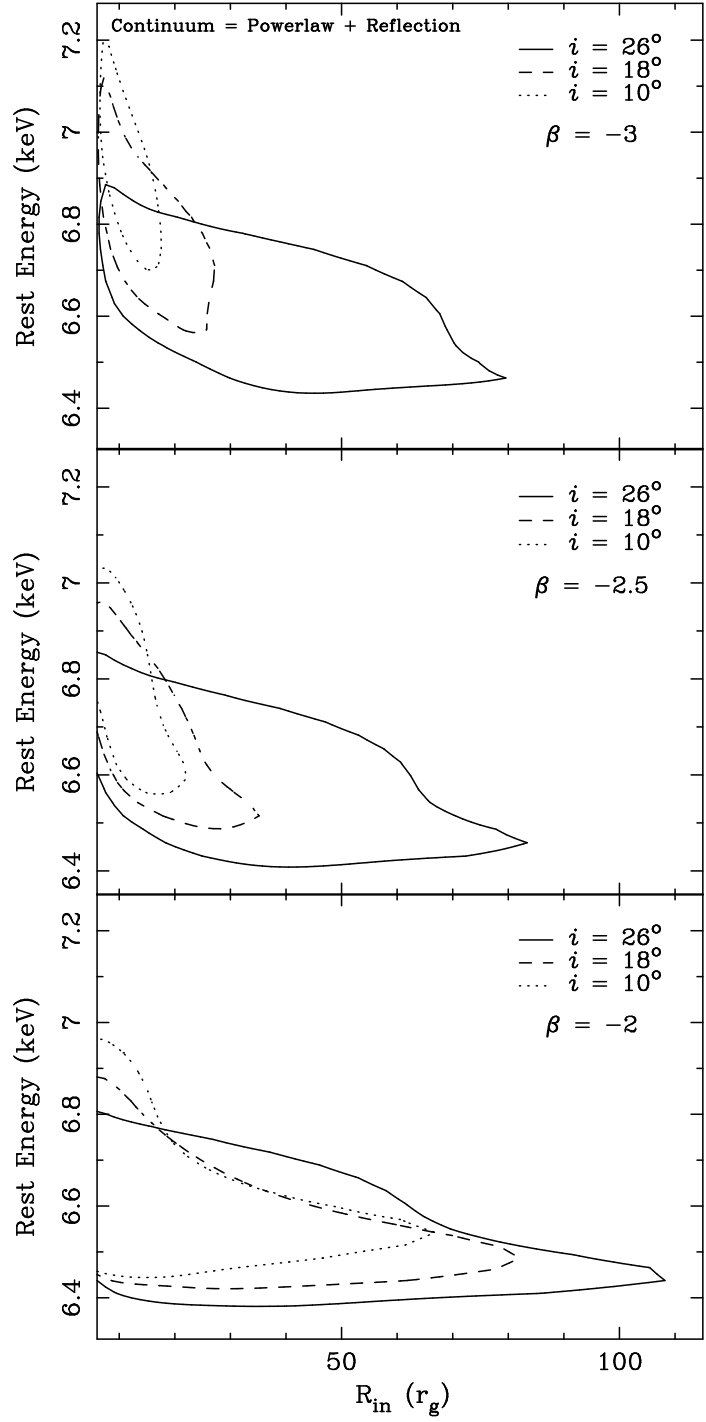}
\caption{\label{diskline2} Confidence contours (90\%) for the diskline
fit to the \fek emission line in $R_{in}$--$E$ space, using the
Powerlaw + Compton Reflection continuum model (Model \#3) 
for $i$ = 26$^{\rm \circ}$ (solid), 18$^{\rm \circ}$ (dashed), and 10$^{\rm \circ}$ (dotted)
and $\beta$ = --2.5.}
\end{figure}

Since the various line model parameters are highly interdependent, we
show the fit results in the form of contours (Figs. \ref{gauss},
\ref{diskline}, and \ref{diskline2}), rather than list them in a table. 
For the Gaussian model, 90\% confidence contours in the line energy
and the FWHM of the line profile and contours in the EW and the FWHM
are shown. For the disk line model, 90\% confidence contours in the
inner radius of the disk ($R_{in}$) and the line energy are
constructed for models with $\beta$ = --2, --2.5 or --3 and $i$ = 10,
18 or 26$^{\rm \circ}$, with $R_{out}$ fixed to be 500 $r_{g}$; the
contours are very similar for $R_{out}$ = 1000 $r_{g}$.

The improvement in the fit with the addition of an emission line,
whether modeled by a Gaussian or a disk line, is significant. For the
Powerlaw + soft Gaussian continuum model $\chi^{2}$ = 314 for 172
d.o.f. when no line is included, whereas $\chi^{2}$ = 163 for 169
d.o.f. for the Gaussian model. For the disk line model, $\chi^{2}$ =
159, 165, and 173 for 169 d.o.f. for $\beta$ = --3, --2.5, and --2
respectively and $\Delta\chi^{2}$ is not very sensitive to the
inclination angle. For the Powerlaw + Reflection continuum continuum,
$\chi^{2}$ = 279 for 172 d.o.f. when no line is included, whereas
$\chi^{2}$ = 184 for 169 d.o.f. for the Gaussian model. For the disk
line model, $\chi^{2}$ = 186, 187, and 190 for 169 d.o.f. for $\beta$
= --3, --2.5, and --2 respectively. The residuals are equally well
modeled by the Gaussian and the disk line, but for the Powerlaw + soft
Gaussian continuum model, a disk line model with $\beta$ = --2 is
disfavored by the data because of the large value of $\chi^{2}$.

As discussed in \S\ref{cont_sec} and seen in Fig. \ref{spectra} the
MOS 1 data show clear residuals above the \fek line region because
those data prefer a smaller $\Gamma$ than the MOS 2 and PCA
data. Because the properties of a weak line are very dependent upon
the continuum fit (see \S 2), we also present the results for a fit in
which a Gaussian line and a simple powerlaw continuum were fitted {\it
simultaneously} only over the region from 3--10 keV. The photon
indices were allowed to vary independently but \nh was fixed to be $8
\times 10^{21}\; {\rm cm^{-2}}$, because it is unconstrained by these high
energy data. The parameters of the Gaussian line were initially
allowed to vary independently. We found that the line parameters
inferred from the MOS~1 ($E_{\rm MOS 1} = 6.3
\pm 0.2$ keV, $\sigma_{\rm MOS 1} < 0.5$ keV, and $EW_{\rm MOS 1} = 50
\pm 40$ eV) and PCA ($E_{\rm PCA} = 6.4^{\;+0.2}_{\;-0.1}$ keV,
$\sigma_{\rm PCA} < 0.5$ keV, and $EW_{\rm PCA} = 50^{\;+40}_{\;-10}$
eV) data were in agreement. The line is not well constrained by the
MOS 2 data, but when $E$ and $\sigma$ are the same as for MOS 1, the
upper limit to $EW$ is 70 eV, which is consistent with the EW measured
for the MOS 1 and PCA data. Thus for the purposes of investigating the
line properties, the line energy, $\sigma$ and $EW$ were forced to be
the same for all three data sets.\footnotemark\footnotetext{Although
the line is not resolved by any individual data set, when all three
are fitted together the line must be broad, as seen in Fig. 5. This is
because the high S/N PCA data require that the line has $EW > 40$ eV,
but for an {\it unresolved} line the upper limit to the MOS 2 $EW$ is
30 eV; to fit all three data sets, the line must be broad.} As with
the Powerlaw + soft Gaussian and Powerlaw + Reflection continuum
models, the addition of a narrow line leads to a significant
improvement of the fit ($\chi^{2}$ = 155 for 220 d.o.f. compared to
210 for 221 d.o.f when the line is not included.)

An \fek emission line is clearly required to fit the data regardless
of the continuum model. The line is resolved in all cases with a large
bulk velocity (FWHM $> 20,000 \; {\rm km\;s^{-1}}$ and $R_{\rm in} <
100 r_{g}$ for the Gaussian and disk line models respectively) which
implies an origin in an accretion disk as opposed to a distant
reprocessor such as the obscuring torus. However the line need not
form in the innermost regions of the accretion disk. The equivalent
width also has considerable uncertainty ($EW = 40$--$110$ eV), but is
much smaller than the {\it total} $EW$ of the narrow + broad \fek
emission line ($EW \sim 200$~eV) observed in those Seyferts with broad
\fek emission lines.

\subsection{Combined Continuum and \fek Emission Models}\label{combo_sec}

In the previous two sections, we verified that reprocessing features,
especially the Compton reflection bump, are much weaker in 3C~111 than
in Seyferts. However, the models used are not physically reasonable;
the \fek emission line is very broad, implying an origin in the inner
accretion disk, but the Compton reflection, which would necessarily
accompany this disk line, is not allowed by the high energy
continuum. In the following sections, we simultaneously fit the
continuum and \fek emission line, with the goals of finding a
self-consistent model for the data and assessing whether either of the
two scenarios presented in \S\ref{intro} are physically
reasonable. The soft Gaussian component is included in all models and
the best-fit parameters are listed in Table \ref{comb_param} and
discussed below.

\subsubsection{Truncated Accretion Disk - Models \#7a,b}\label{refsch}

The total spectrum from 0.4--100 keV is fitted with a model which
includes Compton reflection and an \fek emission line, both of which
arise from the reprocessing of the primary X-ray emission by a neutral
or moderately ionized accretion disk (Model \#7a). We use the {\sc
refsch} routine, which implements the ionized disk model described in
\S\ref{cont_sec} (Model \#4), but convolved with a disk line model to
account for relativistic effects \citep[see, for example][]{f89}. The
\fek line is modeled as a disk emission line with $\beta \equiv -3, i \equiv
26^{\rm \circ}, R_{out} \equiv 500 \; r_{g}$ and the normalization of the
emission line is tied to $\Omega/2\pi$ such that $EW_{\rm Fe \,
K\alpha} \equiv 160 \times \Omega/2\pi$~eV \citep{gf91}, which is only
appropriate for a solar abundance of Fe.  Both the continuum and line
parameters are allowed to vary.

The inner radius of the accretion disk is $R_{in} =
110^{\;+350}_{\;-70} r_{g}$ and the energy of the emission line is
consistent with being emitted by nearly-neutral Fe, thus this model
indicates that both the continuum and \fek emission line could arise
from a truncated accretion disk. However, there are residuals in the
spectrum near 6.4~keV. Using the {\it Chandra} High Energy
Transmission Grating, \citet{YP04} found that the narrow cores of \fek
emission lines in Seyferts have an average energy of $\sim 6.4$~keV
and $\sigma
\sim 0.02$~keV. The inclusion of such a narrow Gaussian, in which $E$
and $\sigma$ are fixed to these values, improves the fit slightly
($\Delta\chi^{2}$ = 8 for one d.o.f., Model \#7b). The equivalent
width of this narrow line is $\sim 25$~eV. \citet*{ghm94} found that
for disk inclinations similar to that of 3C~111, a Compton-thin torus
(with \nh $\sim 10^{23} \; {\rm cm^{-2}}$) will contribute an \fek
emission line with an equivalent width of 30~eV while making a
negligible contribution to the Compton reflection bump. The line can
also originate by transmission through dense clouds along the line of
sight. However, with the addition of the narrow line, the reflection
fraction, and consequently the equivalent width of the emission line
arising from the disk, are very small ($\Omega/2\pi = 0.09
^{\;+0.04}_{\;-0.05}$). Given the weakness of the disk line it is
impossible, with these data, to place any meaningful constraints on
either the energy of the disk emission line or the inner radius of the
accretion disk.

\begin{figure}
\epsscale{0.75}
\rotatebox{-90}{\plotone{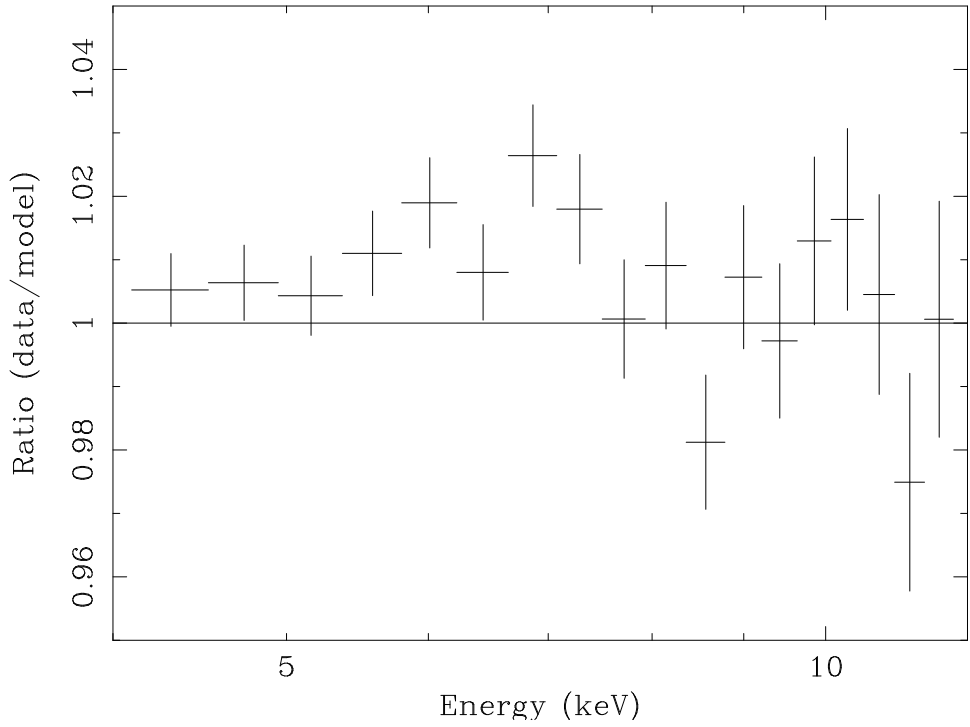}}
\caption{\label{refsch_resid} Ratio of a simulated PCA data set based upon
model \#7a (without the disk emission line component) to the Powerlaw
+ soft Gaussian model (Model \#6). The shape of the residuals is
indicative of curvature in the former continuum model, which can mimic
a broad \fek line.}
\end{figure}

The results of these fits suggest that only a very weak ($EW \sim
10$--20 eV) broad emission line is required and the data are primarily
fitted by a narrow emission line. This is in sharp contrast to the
results of \S\ref{line_sec}, which suggest that the \fek line is
stronger ($EW$ = 40--110 eV) and very broad ($FWHM > 20,000 {\rm
km\;s^{-1}}$).  This is due in part to the fact that there is
significant broad curvature in the {\sc refsch} continuum. Using the
{\sc fakeit} command in {\sc XSPEC} a data set was simulated based
upon the {\sc refsch} model parameters for Model \#7a, but {\it
without} the disk emission line component; when these data are fitted
with the Powerlaw + soft Gaussian model (\#6), there are broad
residuals in the \fek emission line region, as seen in
Fig. \ref{refsch_resid}.  This curvature is likely due to the
relativistically blurred \fek edge, as we noticed that the excess
emission was greater for larger values of $\Omega/2\pi$ but less
noticeable for smaller values of $R_{in}$. If the {\sc diskline}
component of Model \#7a is replaced with a Gaussian, the line only has
$EW = 40 \pm 10$ eV and $\sigma = 0.3^{\;+0.2}_{\;-0.1}$ keV, which is
not inconsistent with the {\it total} $EW$ (narrow + broad) of $\sim
40$ eV inferred from model
\#7b. Thus the broad residuals studied in
\S\ref{line_sec} can be fitted with a combination of continuum
curvature as well as broad {\it and} narrow line emission.

The fit results for this model (\#7b) are {\it consistent} with the
emission arising from the reprocessing of the primary X-ray continuum
by an accretion disk, although a distant reprocessor also contributes
to the flux of the \fek emission line. Because the component of the
line arising from the accretion disk is very weak, $R_{in}$ is not
constrained and the data do not {\it require} the disk to be
truncated. However a larger value of $R_{in}$ (such as found for Model
\#7a) is more {\it physically} attractive because it is consistent
with the small value of $\Omega/2\pi$ found.

\subsubsection{Highly Ionized Accretion Disk - Models \#8a,b,c}\label{rf-pexriv}

As discussed in \S\ref{intro}, the reflection features from a highly
ionized disk are very weak; when fitted with the {\sc pexriv} and {\sc
refsch} routines in {\sc XSPEC}, as we have done thus far, the value
of $\Omega/2\pi$ is misleadingly low \citep*{rfy99}. Therefore, we now
fit the data with the constant density ionized disk model described by
\citet{rf93} and \citet*{bif01}, which is available as a table model for use in
XSPEC (Model \#8a). We refer to this model as {\sc rf-pexriv} to avoid
confusion with the previously discussed {\sc pexriv}
model. \citet{brf02} compared this model to the {\it ASCA} spectrum of
the BLRG 3C~120 and found that the spectrum is well fitted with $\xi$
= $ 4000 {\rm erg\;cm\;s^{-1}}$ and $\Omega/2\pi$ fixed at unity.

The disk is modeled as a slab of gas with a constant density of
10$^{15}$ cm$^{-3}$, which is illuminated by a powerlaw with index
$\Gamma$ and a sharp cut-off at 100 keV, and the ionization parameter
extends beyond 10$^{4}$. (Here the ionizing flux is defined from
0.01--100~keV.)  This model includes the emission from the \fek line,
so the emission line and continuum are fit self-consistently.  The
Comptonization of the reprocessed photons as they emerge from the
dense disk is also included, but gravitational and
inclination-dependent radiative transfer effects are not.

\begin{figure}
\epsscale{0.7}
\rotatebox{-90}{\plotone{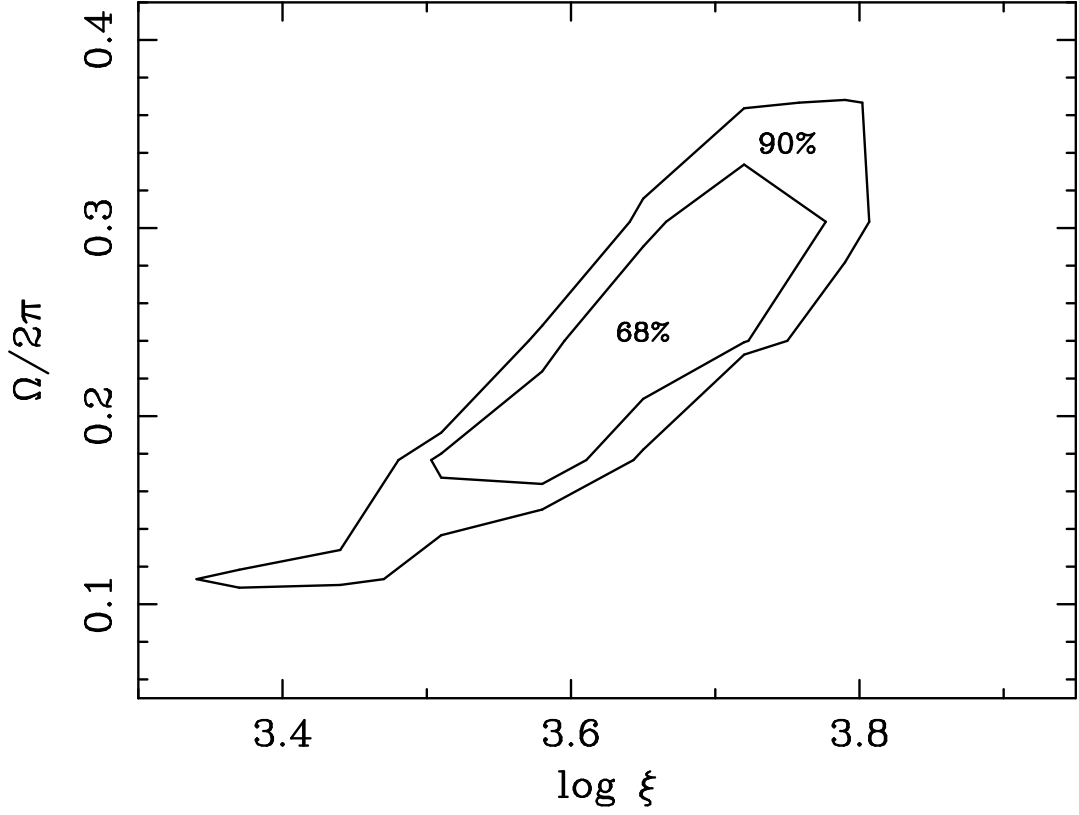}}
\caption{\label{rf_plot} Confidence contours in the log$\;\xi$--$\Omega/2\pi$ plane
for the constant density ionized accretion disk model described in
\S\ref{rf-pexriv}.}
\end{figure}

Even though the best-fit ionization is quite large ($\xi \sim 4000$)
the reflection fraction is $\Omega/2\pi = 0.3 \pm 0.1$; $\Omega/2\pi
\sim 1$, as observed in Seyferts, is not allowed (see Fig. 
\ref{rf_plot}). As before, the addition of an narrow emission line
with $E \equiv 6.4$ keV improves the fit slightly ($\Delta\chi^{2} =
7$ for 1 extra d.o.f; Model \#8b); the $EW$ of the narrow line is
$\sim 15$~eV but the reflection fraction is further reduced to
$\Omega/2\pi = 0.2$.  The results of these fits suggest that even when
the disk is allowed to be highly ionized, $\Omega/2\pi$ is still
rather small. However, if gravitational effects are included (as in \S
5.3.2), larger values of $\Omega/2\pi$ {\it might} be allowed, because
the reflection features would be blurred.

As a test, we have convolved the {\sc rf-pexriv} model with a disk
emission line\footnotemark\footnotetext{The convolution was carried
out using a code kindly provided by A.C. Fabian.} with $\beta \equiv
-3, i \equiv 26^{\rm \circ}, R_{out} \equiv 400 \; r_{g}$ and also
included a narrow (unconvolved) 6.4 keV Gaussian (Model \#8c). The
response functions were extended using the {\sc extend} command in
{\sc XSPEC} and the HEXTE data above 80 keV were ignored because the
model is only defined up to 100 keV. We find that the fit is improved
when the relativistic effects are included but that the best-fit
parameters (see Table 3) are very similar; in particular $\Omega/2\pi
\sim 0.2$ and $EW_{n} \sim 15$ eV. (A model in which $R_{in} \equiv
100 r_{g}$ fits equally well.)  This model is computationally
expensive and it is impossible with the available computer facilities
to extensively explore parameter space for this model or to determine
error bars for the parameters. Instead, we determined best-fit models
for specific values of $\Omega/2\pi$ (0.6 and 1.0) and we find that
these larger values are not absolutely ruled out, but they are not
favored by the data ($\Delta\chi^{2}$ = +7 and +13 for 1
d.o.f. respectively, compared to the best fit model
\#8c).

The models presented in this section provide very good fits to the
data and it is certainly possible that the accretion disk is
ionized. However, the second scenario proposed in \S\ref{intro}, (that
$\Omega/2\pi \sim 1$ and the reprocessing features are weak {\it only}
because the disk is highly ionized), is not {\it preferred} by the
data, although it is not completely ruled out.

\subsubsection{Partial Covering - Model \#9}\label{pcfabs}

Recent observations with {\it XMM} have shown that, contrary to the
early results from {\it ASCA}, many Seyfert 1s possess {\it only} a
narrow \fek line with no evidence for a broad, distorted line
\citep{r03}.  In some of these objects, broad residuals are clearly
present in the \fek region of the spectrum, but they are equally well
fitted by a broad diskline and a model in which the continuum is
partially covered by an absorber with a column density of $10^{22-23}
\; {\rm cm^{-2}}$ (see, e.g. 1H0419-577, Pounds et al. 2004a; NGC
4051, Pounds et al. 2004b).  This heavily absorbed powerlaw spectrum
turns over at an energy near the \fek emission line, thereby
introducing curvature into the continuum spectrum than can mimic a
broad \fek line. Here we test whether the broad residuals in the \fek
region of the spectrum of 3C~111 are simply an artifact of partial
covering.

\begin{figure}
\epsscale{0.7}
\rotatebox{-90}{\plotone{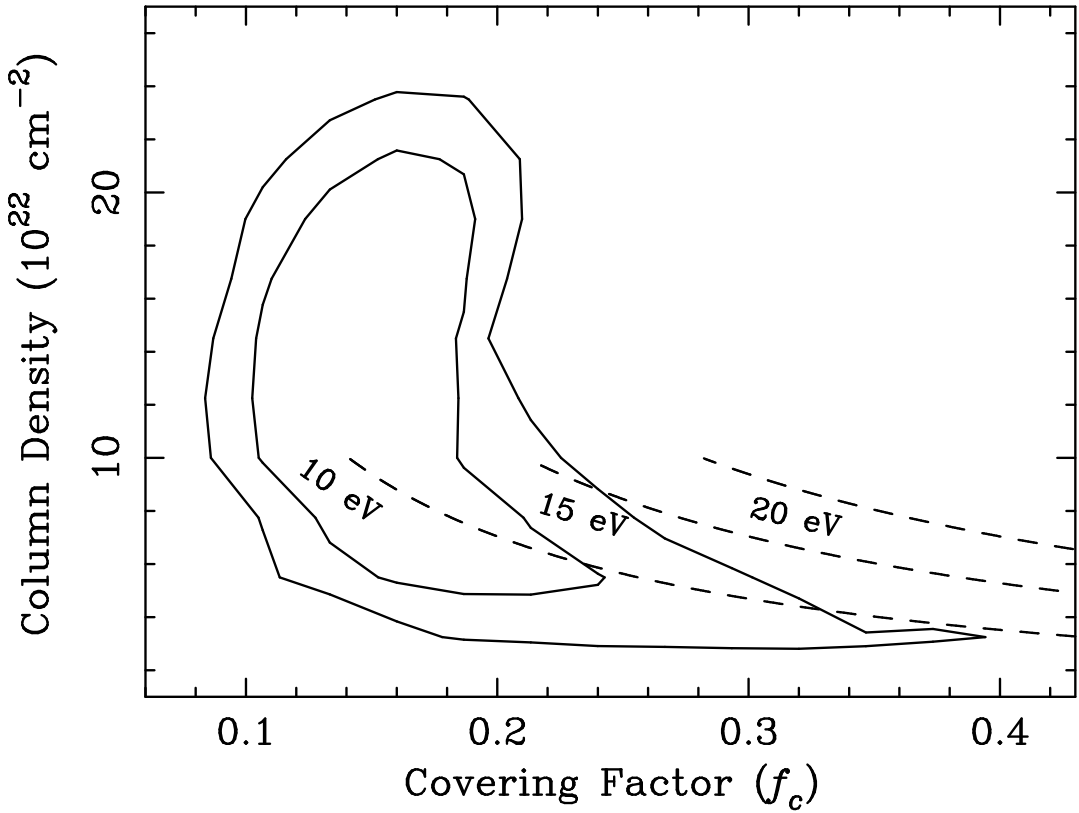}}
\caption{\label{nh2_cfrac} Confidence contours in the column density and 
covering fraction for the partial covering powerlaw model described in
\S\ref{pcfabs}. As described in \S\ref{pcfabs}, the absorber is
expected to produce an \fek emission line. Overlaid on the contours
are lines of constant \fek equivalent width (10,15, and 20 eV) as a
function of $N_{\rm H,2}$ and $f_{c}$ as prescribed by equation (1).}
\end{figure}

\begin{figure}
\epsscale{0.7}
\plotone{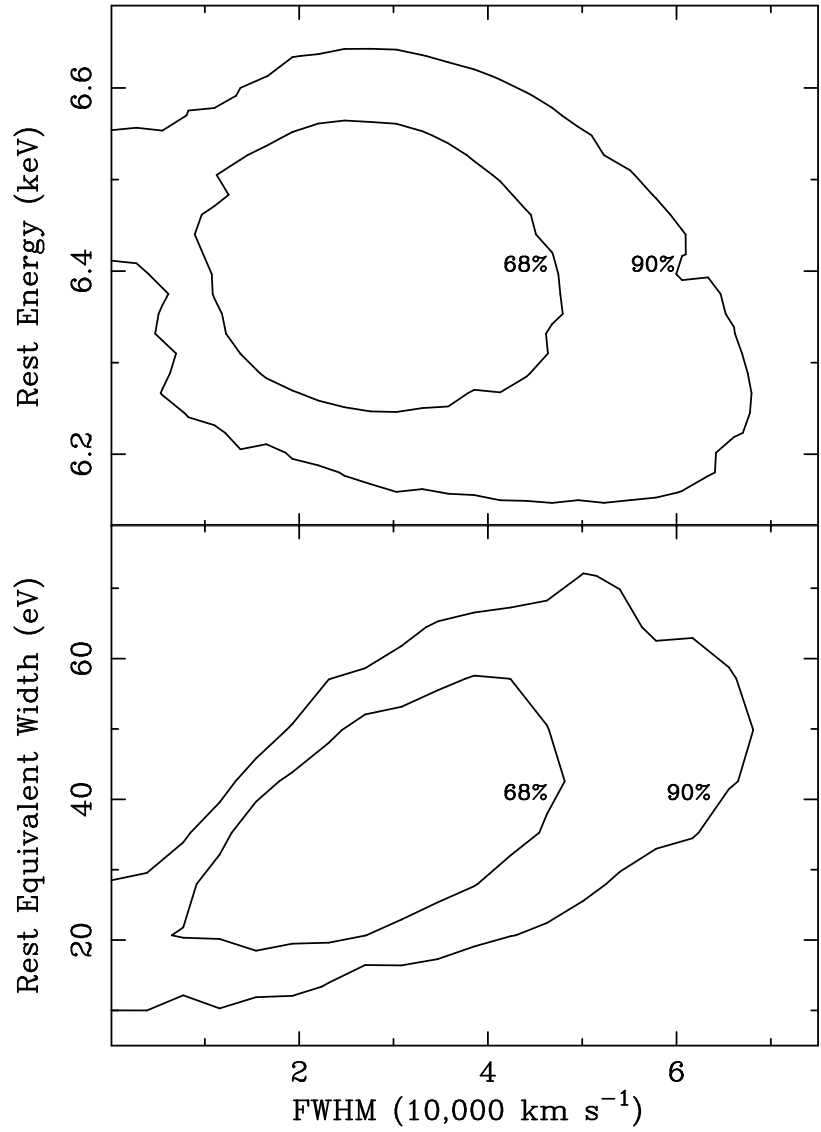}
\caption{\label{pc_gauss} 
Confidence contours for the Gaussian fit to the residual \fek emission
line from the partial covering model (with the soft Gaussian included)
described in \S\ref{pcfabs}. The equivalent width is defined to be
the line flux divided by the specific continuum flux at energy E.}
\end{figure}

The data are fitted with an absorbed powerlaw model that includes an
additional partial covering absorber with column density $N_{\rm H,2}$
and covering factor $f_{c}$, implemented with the {\sc pcfabs}
routine.  There are clear residuals in the spectrum so a Gaussian
emission line is also included (Model \#9). Confidence contours for
$N_{\rm H,2}$ and $f_{c}$ are shown in Fig. \ref{nh2_cfrac}.  The
emission line has an energy of $\sim 6.4$~keV and an $EW \sim 30$~eV
and, as can be seen in Fig. \ref{pc_gauss}, the line is not resolved at
the 90\% confidence level. We find that even when the data from 3--8
keV are excluded from the fit, the same values of $N_{\rm H,2}$ and
$F_{c}$ are found, which suggests that continuum itself is better
fitted by this partial covering model than a simple powerlaw.  We note
that this model, in which {\it none} of the primary emission is
reprocessed by a dense, geometrically thin accretion disk, could be
considered to be an extreme case of scenario \#1 presented in
\S\ref{intro} (i.e. the accretion disk is truncated).

\section{DISCUSSION}\label{discussion}

\subsection{Origin of the Low Energy Component}\label{soft}

Many of the continuum models presented in \S\ref{cont_sec} do not
provide a satisfactory fit to the low energy data. We find that the
inclusion of a low energy component, described either by a $\sim
1$~keV Bremsstrahlung (or $\sim$ 0.3 keV blackbody) or a broad
Gaussian line with $E \sim 1.5$~keV and $\sigma \sim 0.5$~keV, greatly
improves the fit. Although it appears that this component must be
included to obtain a good description of the 3--100~keV spectrum,
it would be reassuring to determine whether there are any plausible
sources of this low-energy component. 

Previous {\it ASCA} and {\it ROSAT} observations have shown no
evidence for a soft component in the spectrum of 3C~111, however the
large and uncertain column density would have made its detection
difficult. Soft components have been observed in other BLRGs.  The
BLRG 3C~382 shows a soft excess which cannot be entirely explained by
the known extended halo of hot gas which surrounds the host galaxy
\citep{g01a}. The BLRG 3C~120 exhibits a clear soft excess, that
contributes 20\% of the 0.6--2~keV flux that \citet{bfi04} model with
a $\sim 0.3$--0.4~keV Bremsstrahlung. However, the origin of these
soft components are not well understood. 

An alternative possibility is that the soft energy component simply
compensates for an error in the calibration of the MOS effective area
curves.  The count rate near 1 keV is very large, with some data bins
containing more than 4000 counts. The statistical uncertainties ($\sim
2\%$) are on the order of the uncertainties in the calibration of the
MOS effective area curves, which can be as large as $\sim 10\%$
(XMM-Newton helpdesk, private communication). For this reason, we also
modeled the low energy component as a Gaussian with $E \sim 1.5$~keV
and $\sigma \sim 0.5$~keV. The peak amplitude of the emission line is
only $\sim 8\%$ of the continuum flux, so it is not impossible that
the soft Gaussian component is compensating for an error in the MOS
effective area curves. The soft Gaussian component is admittedly
ad-hoc. However, this component has been chosen, instead of a thermal
component, primarily because it affects a relatively isolated region
of the spectrum (compared to the Bremsstrahlung model) while at the
same time it produces a better fit to the low energy residuals. There
is independent evidence that there is a known excess in the MOS data,
similar to what are present in these data \citep[see Fig. 11 in
][]{K04}. We have verified that including the Soft Gaussian component
{\it only} in the two MOS data sets does not affect the results
presented here.

\subsection{Interpretation of the Spectral Models}

As seen in \S\ref{cont_sec} and \S\ref{combo_sec}, the high energy
data do not allow for a significant Compton reflection bump, even if
the disk is highly ionized (see \S\ref{rf-pexriv}, Models
\#8a,b,c). Therefore these data disfavor the presence of a standard,
optically-thick, geometrically-thin accretion disk (neutral or
ionized), which extends to small radii, in which case $\Omega/2\pi
\sim 1$.  For comparison, if the outer, thin accretion disk
transitions at some radius to a vertically extended structure, (such
as an ion-torus), then $\Omega/2\pi < 0.3$ \citep{ch89,zls99},
consistent with the values of $\Omega/2\pi$ we have obtained.

Although the results of \S\ref{line_sec} suggest that there is a broad
($FWHM > 20,000 \;{\rm km \; s^{-1}}$ \fek line in 3C~111, these
residuals can be partly attributed to curvature in the continuum,
either due to relativistic blurring of the \fek edge (\S\ref{refsch})
or partial covering by a dense absorber (\S\ref{pcfabs}). Weak \fek
line emission with $EW \sim 40$ eV is still required, but the majority
of the flux in this line most likely arises in a distant
reprocessor. Thus, from the point of view of fitting the spectrum of
3C~111, it is not necessary to invoke reprocessing by an accretion
disk to model the data and there is no obvious reason to prefer the
truncated disk models (\#7a,b) over the more simple partial covering
model (\#9).

However, it is not immediately obvious which structure could be
obscuring the primary X-ray source. The inclination of 3C~111 is less
than $26^{\rm \circ}$ so unless the obscuring torus has a {\it very}
large opening angle or is very extended vertically, it cannot be
responsible for the partial covering. An alternative source of partial
covering is Compton-thin (\nh $< 10^{23} \; {\rm cm^{-2}}$) clouds
along the line of sight. These same clouds could also contribute to
the unresolved \fek emission line. The expected $EW$ of the \fek line
transmitted by such clouds is \citep{h82}

\begin{equation}
EW = \frac{350f_{c}}{\Gamma + 2}\left(\frac{6.4}{7.1}\right)^{\Gamma-1}\left(\frac{N_{H,2}}{10^{23} {\rm cm}^{-2}}\right)\left(\frac{A_{Fe}}{4 \times 10^{-5}}\right) {\rm eV}
\end{equation}

\noindent where $f_{c}$, $N_{H,2}$, and $\Gamma$ are the same covering factor,
column density, and powerlaw index as found from the partial covering
model fit, and $A_{Fe}$ is the Fe mass fraction. The above equation
reduces to $EW$ = $71f_{c}(N_{H}/10^{23} \; {\rm cm}^{-2})$ eV. For a
given $EW$ of the \fek emission line, there is one-to-one relationship
between $N_{H,2}$ and $f_{c}$; tracks for $EW$ = 10, 15, and 20 eV are
shown in Fig. \ref{nh2_cfrac}. As can be seen, the partial covering
absorber can contribute an \fek line with an equivalent width of at
most 15 eV.  However, if the clouds have a super-solar abundance, the
\fek emission would be enhanced.  It is likely that the bulk of the
narrow \fek emission line is formed in the obscuring torus, which can
contribute $\sim 30$ eV to the total equivalent width. Transmission
through clouds with $N_{H,2} > 10^{23} \; {\rm cm}^{-2}$ could also
make a small contribution to the observed \fek line, but it is very
difficult, without detailed modeling, to estimate the $EW$ of an \fek
emission line in this case. 

Although BLRGs, on average, do not have large intrinsic column
densities \citep{sem99}, the BLRG 3C~445, which also has an FR~II
radio morphology, is similar to 3C~111 in several respects. The
intrinsic absorber has a large column density \citep[$N_{H,int} \sim
10^{23} \; {\rm cm}^{-2}$,][]{w98,sem99} but unlike 3C~111, the
absorber in 3C~445 covers the source almost completely \citep[$f_{c}
\sim 0.8$;][]{sem99}. The reflection fraction in 3C~445 is also very 
small $\Omega/2\pi < 0.2$, but there is a very strong \fek emission
line with $EW \sim 150$~eV \citep{w98}; these authors postulate that
the \fek emission line arises in a shell of cold material with \nh
$\sim$ (2--5)$\times 10^{23} \; {\rm cm}^{-2}$ which is isotropically
irradiated from the central source. Thus, 3C~445 and 3C~111 might in
fact be quite similar, with the only difference being the covering
factor of the absorber.

The partial covering absorber model for 3C~111 would be best tested by
UV and soft X-ray observations, because these absorbing clouds should
also exhibit emission lines and absorption edges in these
regimes. With combined UV and X-ray information, it would be possible
to place constraints on the physical conditions and possibly the
location of the absorbing material. Unfortunately, the Galactic column
density is so large ($N_{H,Gal} \sim 10^{22} \; {\rm cm}^{-2}$) that
these observations are not possible. However, high S/N UV and X-ray
observations of other BLRGs, which do not suffer from such a high
Galactic extinction, might reveal that intrinsic absorption is not
uncommon in BLRGs. If the covering factor of the absorber is small, as
inferred for 3C~111, then presence of a partial covering absorber
could easily be overlooked because the resulting continuum curvature
is in the same region as the expected \fek emission line.

Alternatively, the partial covering model presented here might simply
be a parameterization of the truncated accretion disk ({\sc refsch})
model presented in \S\ref{refsch}, in which significant continuum
curvature near the \fek line is present (Fig. \ref{refsch_resid}).
When a Gaussian line was added to that model in place of a disk
emission line, the $EW$ and $\sigma$ are very similar to those
obtained when using the partial covering continuum (see
Fig. \ref{pc_gauss}).  Furthermore, when Seyfert 1s that possess broad
\fek line features are fitted with a partial covering model, one
obtains $N_{\rm H,2} \sim$ 3--4 $\times 10^{23} {\rm cm^{-2}}$ and
$f_{c} < 0.35$ \citep{g03}. As with 3C~111, even when the data from
3--8 keV are excluded from the fit, the values of $N_{\rm H,2}$ and
$f_{c}$ are similar (Gelbord, private communication). These values of
$N_{\rm H,2}$ are somewhat larger than found for 3C~111, but as
discussed in \S\ref{refsch}, the curvature in the {\sc refsch} model
appears to be more extreme for larger values of $\Omega/2\pi$; when a
simulated data set based on a {\sc refsch} model with $\Omega/2\pi =
0.75$ and $R_{in}=6 r_{g}$ (using the {\sc fakeit} command in {\sc
XSPEC}) is fitted with a partial covering model, we find $N_{\rm H,2}
\sim 3\times 10^{23} {\rm cm^{-2}}$. Thus it is possible that, as with
the Seyfert 1s studied by \citet{g03}, the partial covering model for
3C~111 (Model \#9) is simply a parameterization of continuum curvature
and is not due to a real absorber.

\section{CONCLUSIONS}\label{conclusions}

In this paper, we present the results of an analysis of simultaneous
observations of the BLRG 3C~111 with {\it XMM-Newton} and {\it RXTE}.
The flux is moderately variable, but there is no evidence for
significant spectral variability. The continuum emission has been
fitted with a wide variety of models and in all instances there are
unacceptably large residuals at low energies. These are well fitted by
a Gaussian component, which is included in all of the continuum models
and is attributed to uncertainties in the calibration of the MOS
detectors. Clear, broad, residuals are also present near the \fek
emission line region for all models. These can be fitted with a broad
\fek emission line with an equivalent width of $\sim 40$--100~eV,
which is weaker than those observed in Seyferts 1s. The high energy
{\it RXTE} data strongly disfavor the presence of a strong Compton
reflection bump, even if the disk is highly ionized. This result gives
strong support to the hypothesis that the geometry of the accretion
flow in BLRGs is different from that in Seyfert 1 galaxies.

We find that continuum curvature is a primary source of the broad
residuals seen in the \fek line region, although line emission is also
required.  The data are consistent with a model in which a weak
Compton reflection bump and \fek disk emission line are formed in a
truncated accretion disk which transitions to a vertically extended
structure (such as an ion torus) at small radii.  A less complex
model, in which the primary X-ray source is partially covered by a
dense ($10^{23} \; {\rm cm^{-2}}$) absorber, also provides a
satisfactory fit to the data. In both models, an \fek emission line
that most likely arises in a distant reprocessor, such as a
Compton-thin obscuring torus or dense clouds along the line of sight,
is also present. Given these data, it is not possible to distinguish
between these two models. However, it is very likely that the partial
coverage model is simply a parameterization of the more complicated
model in which the primary emission is reprocessed by an truncated
accretion disk. Moreover, the partial coverage model is unappealing in
view of the small inclination angle of the jet in 3C~111.

\acknowledgments

This research was funded by NASA Grant NAG5-9982. K.T.L acknowledges
support from a fellowship granted by the NASA Graduate Student
Research Program (NGT5-50387) and the Pennsylvania Space Grant
Consortium.  We would like to thank the XMM helpdesk, particularly
Matteo Guinazzi, for invaluable assistance in determining the source
of the calibration errors in the p-n data.  The code to perform the
relativistic blurring of the ionized disk model (\S\ref{rf-pexriv})
was generously provided by A.C. Fabian and R. Johnstone. Finally, we
are grateful to J. Gelbord, D. Ballantyne, and the anonymous referee
for their many useful comments and suggestions.

\begin{appendix}
\section{Inclination Angle}\label{inclination}

Following the methods of \citet{ehl96}, we use the observed
superluminal motion in the radio jet of 3C~111 and the projected
linear size of the radio lobes to place constraints upon the
inclination angle of the jet, and thus the accretion disk.

The measured proper motion, $\mu$, in the jet is $1.54 \pm 0.2 {\rm \;
arcsec \;yr^{-1}}$ \citep{vc94}. The apparent velocity, relative to
the speed of light, is given by $\beta_{app} \simeq
47.4\;\mu\;z\;h^{-1}$, where $z$ is the redshift, and $h$ is Hubble's
constant, in units of 100~${\rm km \; s^{-1} \; Mpc^{-1}}$.  Thus, for
3C~111 $\beta_{app}$ = 5.1 $\pm$ 0.7, assuming $h$ = 0.7, which
implies either $i < 13^{\rm \circ}$ or $10^{\rm \circ} < i < 26 ^{\rm
\circ}$. We note that \citet{eh98},
\citet{r98} and \citet{esm00} neglected the factor of $h$ in the denominator, 
and thus underestimated $\beta_{app}$ and overestimated the upper
limit on the inclination angle $i$.

A lower limit on the inclination can be inferred from the size of the
radio lobes. \citet{n93} measure the largest angular size of the radio
lobes to be 275$''$, which translates to 184.8 kpc/$h$. \citet{gp78}
find that for powerful radio sources (log $P_{1215\;MHz} > 25.0$), the
radio lobes have an average intrinsic size of 180 kpc$\; h^{-1}$, with
a tail of sources which extends to 500 kpc$\; h^{-1}$. Assuming the
radio lobes of 3C~111 have an intrinsic size of 500 kpc$\; h^{-1}$,
the inclination angle must be greater than 21.7$^{\rm \circ}$.
However, it is possible that 3C~111 is a giant radio galaxy (GRG) and
the intrinsic size of its radio lobes could be much larger. There are
some enormous giant radio galaxies, such as 3C~236 (3 Mpc$\; h^{-1}$;
Nilsson et al. 1993) and NVSS 2146+82 (2 Mpc$\; h^{-1}$; Palma et
al. 2000), but these sources are rare and the size distribution of
GRGs drops off rapidly for sizes greater than 1 Mpc$\; h^{-1}$
\citep{s01}. Assuming 3C~111 is not larger than 1 Mpc$\; h^{-1}$, we
have $i >10.6^{\rm \circ}$. Thus we adopt a conservative range in the
inclination angle, $10^{\rm \circ} < i < 26^{\rm \circ}$, keeping in
mind that unless 3C~111 is a GRG, it is very likely that $21^{\rm
\circ} < i < 26^{\rm \circ}$.

\end{appendix}

%Observation Log Table
\begin{deluxetable}{lccl}
%\tabletypesize{\footnotesize}
\tablecaption{\label{exposures}Observation Log}
\tablewidth{6.0 in}
\tablehead{
   \colhead{Instrument}  & {Exposure Time (ks)} & {Exposure Time (ks)} & {Count Rate (counts s$^{-1}$)\tablenotemark{a}}\\
   \colhead{} & \colhead{} & \colhead{post-filtering} & \colhead{post-filtering}
          }
\startdata
\multicolumn{4}{c}{{\it XMM} - Observation Date: 2001 Mar 14 13:03 -- 2001 Mar 15 01:16}\\
pn     & 42.3 & 23.4  & 10.89 \\
MOS 1  & 44.0 & 27.8  & 3.22 \\
MOS 2  & 44.0 & 28.4  & 3.22 \\
RGS 1  & 44.6 & 43.5  & 0.08, 0.05 \tablenotemark{b} \\
RGS 2  & 44.6 & 42.5  & 0.10, 0.04 \tablenotemark{b} \\
\multicolumn{4}{c}{{\it RXTE} - Observation Date: 2001 Mar 14 08:49 -- 2001 Mar 17 03:36} \\
PCA     &       & 56.6  & 6.57 \\
HEXTE 0 &       & 18.2  & 1.14 \\
HEXTE 1 &       & 17.8  & 0.86 \\
\enddata
\tablenotetext{a}{The count rate was measured for the same interval used for fitting. (For the pn data, 0.5--10~keV was used.)} 
\tablenotetext{b}{The 1st and the 2nd order count rates, respectively.}
\end{deluxetable}
%End Table of Observations

%Continuum Fit Parameters Table
\begin{deluxetable}{cccc}
\tabletypesize{\footnotesize}
\tablecaption{\label{cont_param}Best Fit Continuum Parameters}
\tablewidth{7.0 in}
\tablehead{
  \colhead{Model} & \colhead{Model Number} & \colhead{Model Parameters \tablenotemark{a,b}} & \colhead{$\chi^{2}$/d.o.f.}
          }
\startdata
Powerlaw              & 1 & \nh = ($8.1 \pm 0.1$) $\times 10^{21}\; {\rm cm}^{-2}$           & 1005/895\\
                      &  & $\Gamma$ = $1.72 \pm 0.01$                         & \\
\\
Broken Powerlaw       & 2 & \nh = ($8.7^{\;+0.1}_{\;-0.2}$) $\times 10^{21} \; {\rm cm}^{-2}$ &  928/893\\
                      &  & $\Gamma_{1}$ = $1.86^{\;+0.04}_{\;-0.03}$         &\\
                      &  & $E_{\rm{break}}$ = $2.9 \pm 0.2$~keV                &\\
                      &  & $\Gamma_{2}$ = $1.69 \pm 0.01$                     &\\
\\
Powerlaw + Compton    & 3 & \nh = ($8.1 \pm 0.1$) $\times 10^{21} \; {\rm cm}^{-2} $           & 999/892\\
Reflection (neutral)  &  & $\Gamma$ = $1.74^{\;+0.01}_{\;-0.02}$             &\\
\citep{mz95}          &  & $E_{\rm{fold}} > 440$~keV                          &\\
                      &  & $\Omega/2\pi$ = $0.13^{\;+0.06}_{\;-0.07}$        &\\
\\
Powerlaw + Compton    & 4 & \nh = ($8.1 \pm 0.01$)$ \times 10^{21}\; {\rm cm}^{-2}$         & 997/890\\ 
Reflection (ionized)  &  & $\Gamma$ = $1.74^{\;+0.02}_{\;-0.01}$              &\\
\citep{mz95,d92}      &  & $E_{\rm{fold}} > 440$~keV                          &\\
                      &  & $\Omega/2\pi$ = $0.13 \pm 0.06$                    &\\
                      &  & $\xi < 20\;{\rm erg \;cm\;s^{-1}}$                    &\\
\\
Powerlaw + Bremsstrahlung \tablenotemark{c} & 5 & \nh = ($8.5 ^{\;+0.01}_{\;-0.02}$)$ \times 10^{21}\;{\rm cm}^{-2}$ & 938/892\\
                      &  & $\Gamma$ = $1.68 ^{\;+0.02}_{\;-0.01}$             &\\
                      &  & kT = $0.9 \pm 0.2$~keV                             &\\
\\
Powerlaw + Gaussian \tablenotemark{d} & 6 & \nh = ($8.0^{\;+0.1}_{\;-0.2}$)$ \times 10^{21}\;{\rm cm}^{-2}$ & 916/892\\
                      & & $\Gamma$ = $1.69 \pm 0.01$                         & \\
                      &  & $E$ = $1.5^{\;+0.1}_{\;-0.2}$~keV                    & \\
                      &  & $\sigma$ = $0.5 \pm 0.1$~keV                       & \\
\enddata
\tablenotetext{a}{The error bars are the 90\% confidence interval for
1 degree of freedom (i.e. $\Delta\chi^{2} = 2.706$).}
\tablenotetext{b}{The observed 2--10 keV flux, as measured with the MOS 1 and 2
cameras is ($5.8 \pm 0.1$)$\times 10^{-11} \; {\rm erg \; s^{-1}
cm^{-2}}$. The unabsorbed luminosity, at the distance of 3C~111, is
$3 \times 10^{44} {\rm erg \; s^{-1}}$.}
\tablenotetext{c}{The Bremsstrahlung component has a 0.5--2~keV
unabsorbed flux of ($5 \pm 2$)$\times 10^{-12} \; {\rm erg\; s^{-1}\;
cm^{-2}}$ and contributes 10\% of the flux in the 0.5--2~keV band and
1\% of the flux in the 2--10 keV band.}
\tablenotetext{d}{The unabsorbed flux of the emission line 
$1.4^{\;+1}_{\;-0.4} \times 10^{-12} \; {\rm erg\;cm^{-2}\;s^{-1}}$.}

\end{deluxetable}

%Combined Fit Parameters Table
\begin{deluxetable}{cccc}
\tabletypesize{\scriptsize}
\tablecaption{\label{comb_param}Best Fit Parameters for Combined Continuum and \fek Line Models}
\tablewidth{7.0 in}
\tablehead{
  \colhead{Model \tablenotemark{a}} & \colhead{Model \#} &\colhead{Model Parameters \tablenotemark{b}} & \colhead{$\chi^{2}$/d.o.f.}
          }
\startdata
Powerlaw + Compton                & 7a (see \S\ref{refsch}) & \nh = ($7.9 \pm 0.01$)$\times 10^{21}\; {\rm cm}^{-2}$; $\Gamma$ = $1.66 \pm 0.02$                     & 1039/994\\
Reflection ({\tt refsch})         &   & $E_{\rm{fold}} < 120$~keV \tablenotemark{e}; $\Omega/2\pi$ = $0.20^{\;+0.04}_{\;-0.06}$; $\xi < 4\times10^{2}\;{\rm erg\;cm\;s^{-1}}$     & \\
+ Diskline \tablenotemark{c}      &   & $R_{in}$ = $110^{\;+350}_{\;-70} r_{g}$; $E_{d}$ = $6.4 \pm 0.1$~keV; $EW_{d}$ = $32^{\;+6}_{\;-10}$~eV  & \\
\\
Powerlaw + Compton                & 7b (see \S\ref{refsch}) & \nh = ($7.8^{\;+0.2}_{\;-0.1}$)$\times 10^{21} \; {\rm cm}^{-2} $; $\Gamma$ = $1.63^{\;+0.03}_{\;-0.02}$ & 1031/993\\
Reflection ({\tt refsch})         &   & $E_{\rm{fold}} < 150$~keV; $\Omega/2\pi$ = $0.09^{\;+0.05}_{\;-0.04}$; $\xi < 5\times10^{3}\;{\rm erg\;cm\;s^{-1}}$                       & \\
+ Diskline \tablenotemark{c}      &   & $R_{in} < 500 \, r_{g}$; $E_{d}$ = $5.8^{\;+0.8}_{\;-0.2}$; $EW_{d}$ = $14^{\;+6}_{\;-8}$                 & \\
+ Narrow Gaussian \tablenotemark{d}&  & $EW_{n}$ = $25^{\;+8}_{\;-11}$~eV                                          &\\   
\\
Powerlaw + Compton                & 8a (see \S\ref{rf-pexriv}) & \nh = ($8.3 \pm 0.1$)$\times 10^{21} \; {\rm cm}^{-2}$; $\Gamma$ = $1.64^{\;+0.01}_{\;-0.02}$  & 1050/997 \\
Reflection ({\tt rf-pexriv})      &   & $\Omega/2\pi$ = $0.3 \pm 0.1$; $\xi$ = $4^{\;+4}_{\;-2}\times10^{3}\;{\rm erg\;cm\;s^{-1}}$                                                 & \\                    
\\
Powerlaw + Compton                  & 8b (see \S\ref{rf-pexriv}) & \nh = ($8.2 \pm 0.1$)$\times 10^{21} \; {\rm cm}^{-2}$; $\Gamma$ = $1.66^{\;+0.01}_{\;-0.02}$  & 1043/996 \\
Reflection ({\tt rf-pexriv})        &    & $\Omega/2\pi$ = $0.17 \pm 0.06$;a $\xi$ = $4^{\;+4}_{\;-2}\times10^{3}\;{\rm erg\;cm\;s^{-1}}$                & \\
+ Narrow Gaussian \tablenotemark{d} &    & $EW_{n}$ = $15^{\;+12}_{\;-8}$~eV                                                                       & \\
\\
Blurred Powerlaw + Compton          & 8c (see \S\ref{rf-pexriv}) & \nh = $8.2 \times 10^{21} \; {\rm cm}^{-2}$; $\Gamma$ = $1.65$  & 1032/988 \\
Reflection ({\tt rf-pexriv})\tablenotemark{g}   &    & $\Omega/2\pi = 0.2$; $\xi= 3.5\times10^{3}\;{\rm erg\;cm\;s^{-1}}$  & \\
+ Narrow Gaussian \tablenotemark{d} &           & $EW_{n} = 15$~eV                                                                       & \\
\\
Powerlaw                          & 9 (see \S\ref{pcfabs})& \nh = ($8.2^{\;+0.1}_{\;-0.4}$)$\times 10^{21}\; {\rm cm}^{-2}$; $\Gamma$ = $1.76^{\;+0.02}_{\;-0.04}$        & 1000/994\\
+ Partial Covering                &    & $N_{\rm H,2}$ = ($1.5^{\;+0.7}_{\;-1.4}$)$\times 10^{23}\; {\rm cm}^{-2}$; $f_{c}$ = $0.15^{\;+0.07}_{\;-0.05}$ & \\
+ Gaussian                        &    & $E$ = $6.4 \pm 0.2$~keV; $\sigma$ = $0.2^{\;+0.2}_{\;-0.1}$~keV; $EW$ = $30 \pm 20$~eV                           & \\
\enddata
\tablenotetext{a}{The soft Gaussian component (see \S\ref{cont_sec}, Model \#6) is included in all of these models. The
fit parameters were allowed to vary, but the best-fit parameters are consistent with those found for Model \#6: $E \sim 1.5$~keV; $\sigma$ = 0.5~keV;
 and an unabsorbed flux of $\sim\,10^{-12} {\rm erg\;s^{-1}\;{cm^2}}$.}
\tablenotetext{b}{The error bars correspond to the 90\% confidence interval for
1 degree of freedom (i.e. $\Delta\chi^{2} = 2.706$).}
\tablenotetext{c}{The diskline model has $\beta \equiv$ -3, $i \equiv 26^{\rm \circ}$, 
and $R_{out} \equiv 500 r_{g}$. The normalization of the emission line
has been tied to $\Omega/2\pi$, such that $EW \equiv 160 \times
\Omega/2\pi$ eV. The energy and equivalent width of the diskline are
denoted by $E_{d}$ and $EW_{d}$.}
\tablenotetext{d}{The narrow Gaussian has a fixed energy of 6.4 keV and width 
of 0.02 keV. The equivalent width of the narrow line is denoted by $EW_{n}$.}
\tablenotetext{f}{The folding energy was restricted to be greater than 100~keV.}
\tablenotetext{g}{The model was convolved with a diskline model with $\beta \equiv$ -3, 
$i \equiv 26^{\rm \circ}$, $R_{in} \equiv 500 r_{g}$ and $R_{out} \equiv 400 r_{g}$.}
\end{deluxetable}

%Begin lightcurve
%End lightcurve

%Begin Spectrum Figure
%End Spectrum Figure

%Begin NH-Gamma Figure
%End NH-Gamma Figure

%Begin Reflection Continuum Figure
%End Reflection Continuum Figure

%Begin Gaussian Figure
%End Gaussian Figure

%Begin Diskline Figure
%End Diskline Figure

%Begin diskline comparison Figure
%End Diskline Figure

%Begin Refsch resid Figure
%End Refsch resid Figure

%Begin Ross & Fabian figure
%End Ross and Fabian figure

%Begin nh_covering fraction figure
%End nh_covering factor figure

%Begin pcfabs Gaussian figure
%End pcfabs Gaussian figure

\end{document}